\documentclass[aps,letter,prx,reprint,superscriptaddress,notitlepage,floatfix,nolongbibliography]{revtex4-2}

\usepackage{amsmath}
\usepackage{amssymb}
\usepackage{graphicx}
\usepackage{bm}
\usepackage{xcolor}
\usepackage[T1]{fontenc}
\usepackage[utf8]{inputenc}
\usepackage{xpatch}

\makeatletter
\xpatchcmd{\@ssect@ltx}{\@xsect}{\protected@edef\@currentlabelname{#8}\@xsect}{}{}
\xpatchcmd{\@sect@ltx}{\@xsect}{\protected@edef\@currentlabelname{#8}\@xsect}{}{}
\makeatother

\usepackage[colorlinks=false,hidelinks]{hyperref}
\usepackage{bookmark}

\newcommand{\diff}{\mathop{}\!\mathrm{d}}

\begin{document}

\title{Spontaneous brain-wide rotational dynamics constrain cue-evoked behavior}
\author{Sihan Chen}
\affiliation{Leinweber Institute for Theoretical Physics, The University of Chicago, Chicago, IL 60637, USA}
\affiliation{James Franck Institute, The University of Chicago, Chicago, IL 60637, USA}
\author{Yunlong Xu}
\affiliation{Department of Neurobiology, The University of Chicago, Chicago, IL 60637, USA}
\author{Doruk Efe G\"okmen}
\affiliation{NSF-Simons National Institute for Theory and Mathematics in Biology, 186 E. Chestnut Street, Chicago, IL 60611, USA}
\affiliation{James Franck Institute, The University of Chicago, Chicago, IL 60637, USA}
\author{Michel Fruchart}
\affiliation{Gulliver, ESPCI Paris, Universit\'e PSL, CNRS, 75005 Paris, France}
\author{Brent Doiron}
\email{bdoiron@uchicago.edu}
\affiliation{Department of Neurobiology, The University of Chicago, Chicago, IL 60637, USA}
\affiliation{Neuroscience Institute, The University of Chicago, Chicago, IL 60637, USA}
\author{Vincenzo Vitelli}
\email{vitelli@uchicago.edu}
\affiliation{Leinweber Institute for Theoretical Physics, The University of Chicago, Chicago, IL 60637, USA}
\affiliation{James Franck Institute, The University of Chicago, Chicago, IL 60637, USA}
\date{\today}

\begin{abstract}
Uncovering low-dimensional
representations of brain-wide neural dynamics and their relationship to 
behavior remains a central challenge. Here, we show that spontaneous brain-wide rotational dynamics constrain cue-evoked behavior and learning. To systematically establish these findings, we developed a dimensionality-reduction technique called Rotational Component Analysis (RCA) that optimizes the detection of low-variance rotational modes even within individual trials.
We applied RCA to spike train data from motor cortices of monkeys and humans, mouse cortex-wide calcium imaging
and in human functional magnetic resonance imaging (fMRI) and magnetoencephalography
(MEG).
In mice, a rotational mode identified from 
activity recorded at rest predicts cue-evoked behavior: the phase of rotation when a randomly timed cue arrives predicts
whether the animal subsequently pulls the lever. We explain this surprisingly strong correlation using
both a recurrent neural network (RNN) and a minimal two-dimensional model of response
learning, in which the intrinsic rotational mode serves as a dynamical constraint on
response and learning. We further identify double-rotation modes containing two centers
of rotation that separate preparatory from task dynamics in both mouse and human data.
Remarkably, corresponding modes also organize neural dynamics during a non-motor human
task, and rotational
irreversibility captures early psychosis even when a principal component analysis shows
no group difference. Our work establishes rotational dynamics as low-dimensional
coordinates that constrain how ongoing brain states are converted into behavior. More broadly, RCA provides a versatile tool for detecting hidden nonreciprocal dynamics in biological and physical datasets.
\end{abstract}
\maketitle


Although neural activity is high-dimensional, the behaviorally relevant part can often be captured by low-dimensional latent dynamics~\cite{Perich2025,yu2009gaussian,mante2013context,harvey2012choice,benisty2024rapid,macdowell2025multiplexed}. 
Rotation is a recurrent motif in such low-dimensional neural dynamics.
In this context, rotation arises in cyclic activities produced by central pattern generators and rhythmic motor control~\cite{marder1996principles}, but also in non-cyclic actions such as individual
reaches~\cite{churchland2012neural,pandarinath2015neural,michaels2016neural,
lara2018different,russo2018motor,sabatini2024reach}. 
At the brain-wide scale, oscillations are believed to play a central role in cognitive processes \cite{Ward2003,Basar2001,Schnitzler2005}. 
Similarly, rotating and spiral waves propagating through physical brain space have been suggested to coordinate long-range information flows~\cite{xu2023spiral,ye2026rotating}.
These elements suggest rotation as a potential organizing principle for brain-wide neural dynamics. 

Rotations in
ongoing spontaneous neural activity have received far less attention. 
Spontaneous activity refers to intrinsic neural activity that is not a direct response to external cues \cite{Liu2022,tsodyks1999linking,srinath2026structure,berkes2011spontaneous,goldberg2004patterns,arieli1996dynamics}.
By definition, it cannot be temporally aligned with known causes, making it more challenging to study.
Whether rotational dynamics can be robustly identified in spontaneous activity,
and if so, whether they can be linked to future behavior, remain open questions. 

To address these questions, we introduce rotational component analysis (RCA), an
unsupervised dimensionality-reduction method that explicitly prioritizes rotational
dynamics. RCA quantifies rotation through the breaking of mirror symmetry in the
distribution of neural trajectories and searches for a low-dimensional projection that
maximizes a measure of this asymmetry, rather than maximizing explained
variance or reconstruction of the full neural activity~\cite{crooks1999entropy,
seifert2008stochastic,lynn2021broken,gingrich2016dissipation,li2019quantifying}.
Because of its prioritization, RCA can
reveal low-variance rotational modes that may be underrepresented by other 
methods. This feature is particularly important for spontaneous activity, in which
uninstructed behavior and other ongoing processes generate large rotation-irrelevant
fluctuations that cannot be suppressed by trial averaging. RCA can therefore be applied
directly to continuous recordings or pooled individual trials, without event alignment,
trial averaging or behavioral labels.

Here, we
first benchmark RCA using human and macaque motor-cortical spiking activity
during reaching, a setting in which rotational population dynamics are well established. We then apply RCA to mouse cortex-wide
calcium imaging, human functional magnetic resonance imaging (fMRI) and human
magnetoencephalography (MEG), and succeed in detecting spontaneous rotational dynamics in all three
datasets.  In mice, the rotational mode identified during rest exhibits an unusually
strong association with cue-evoked behavior during task sessions. We propose that the
intrinsic rotation serves as a dynamical constraint on both response and learning: cue-evoked responses are more likely, and more readily learned, when they follow the pre-existing rotational flow. 
We reproduce this relationship in a recurrent neural network (RNN) and a minimal
two-dimensional model. In human fMRI, we find that rotational dynamics are weaker in participants with early psychosis than in controls. We further find that double-rotation modes, defined by circulation
around two distinct centers, separate preparatory and task dynamics in both mouse and
human data. Altogether, our results suggest that rotation provides a brain-wide dynamical
coordinate associated with both behavior and learning.

\section*{Results}

\subsection*{APR and RCA identify rotational irreversibility in reduced phase space}
Our goal is to develop a dimensionality-reduction method that projects
high-dimensional neural activity onto the latent plane exhibiting the most pronounced
directed rotation.
For a high-dimensional time series $\bm y(t)$, consider a candidate plane
defined by two orthonormal projection vectors $\bm w_1$ and $\bm w_2$, with coordinates
$x_1=\bm w_1\cdot\bm y$ and $x_2=\bm w_2\cdot\bm y$. We quantify circulation in this
plane using the centered angular current
\begin{equation}
j=(x_1-c_1)\dot x_2-(x_2-c_2)\dot x_1,
\label{eq:angular-current}
\end{equation}
where $\bm c=[c_1,c_2]$ is the empirical center. This quantity is the planar kinematic analogue
of angular momentum, familiar from the orbital motion of a planet around the Sun: its
sign distinguishes clockwise from counterclockwise circulation.

\begin{figure*}[!t]
\centering
\includegraphics[width=0.96\textwidth]{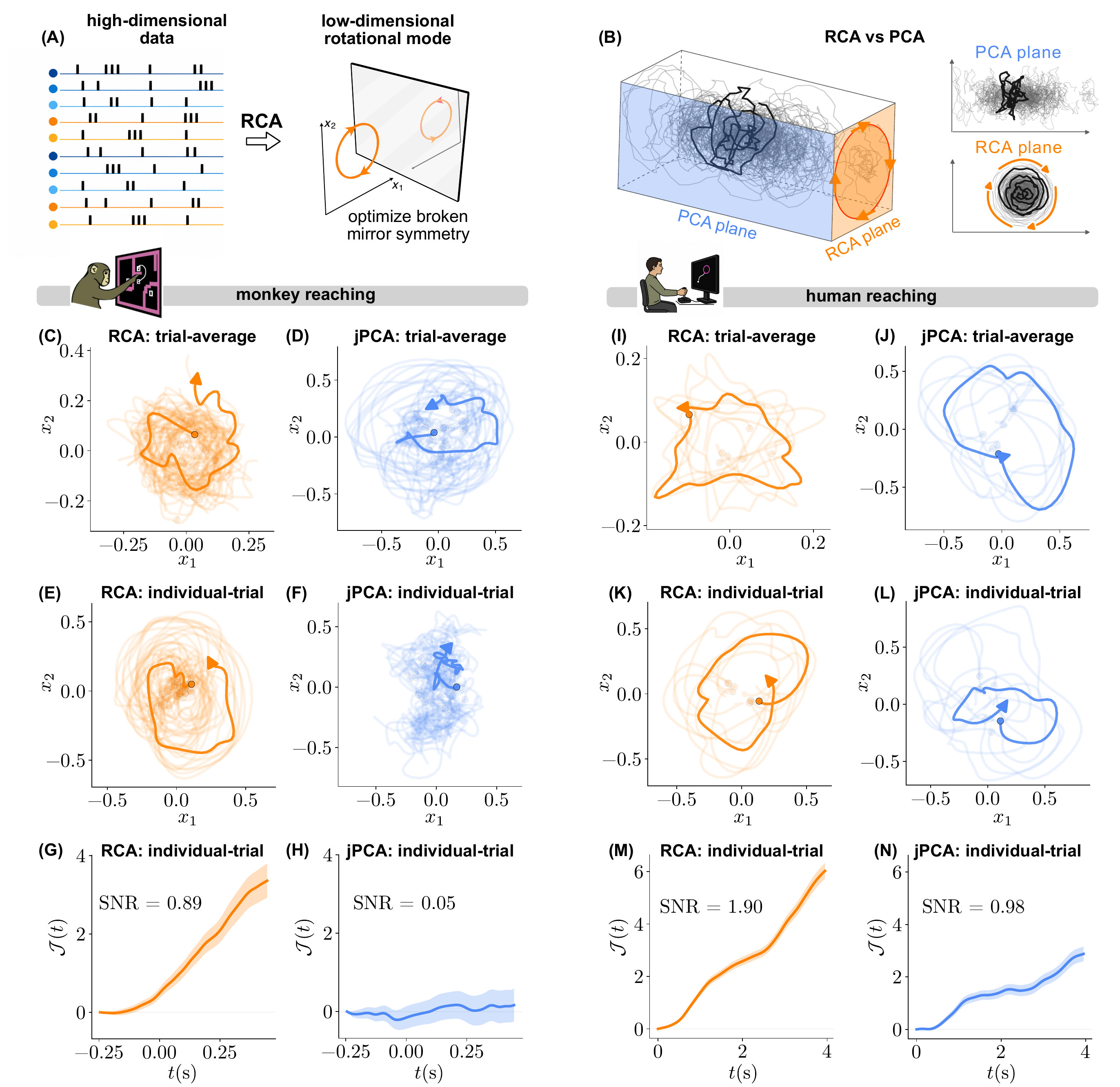}
\caption{\textbf{RCA reveals rotational dynamics in reaching-related spiking activity.}
\textbf{A}, Schematic of rotational component analysis (RCA). High-dimensional data (e.g., spike-train activity) are projected onto a two-dimensional plane
optimized for mirror-symmetry
breaking and directed circulation. \textbf{B}, Schematic comparison with principal
component analysis (PCA). A low-variance rotational plane (orange) can be missed when
the direction of greatest variance is orthogonal to that plane (PC1; blue). The
corresponding PCA and RCA projections are shown at right. \textbf{C--H}, Macaque
motor-cortical activity during delayed reaching. Condition-averaged trajectories are
shown in RCA (orange) and jPCA (blue) planes fitted either to condition-averaged data
(\textbf{C,D}) or to individual trials pooled across reach conditions
(\textbf{E,F}). Pale curves show all reach conditions, and the bold curve highlights
representative maze condition 1. \textbf{G,H}, Normalized cumulative angular current
$\mathcal{J}(t)$ across individual trials from maze condition 1 in the single-trial-fit
RCA (\textbf{G}) and jPCA (\textbf{H}) planes. Lines show the across-trial mean, and
shading denotes $\pm1$ standard error of the mean (s.e.m.). \textbf{I--N}, Corresponding
analyses of human motor-cortical activity during reaching. Condition-averaged
trajectories are shown in planes fitted to condition-averaged data (\textbf{I,J}) or
pooled individual trials (\textbf{K,L}); pale curves show all eight reach conditions,
and the bold curve highlights the representative bottom-target condition.
\textbf{M,N}, Trial-level cumulative angular current for the bottom-target condition in
the individual-trial-fit RCA (\textbf{M}) and jPCA (\textbf{N}) planes. In all trajectory
panels, circles mark trajectory onset and arrowheads indicate temporal direction. The
trial-to-trial signal-to-noise ratio (SNR) in \textbf{G,H,M,N} is the absolute mean
final current divided by its across-trial standard deviation.}
\label{fig:rca-overview}
\vspace{0.9cm}
\end{figure*}

Angular current connects rotation to broken mirror symmetry. Dynamics are
mirror-symmetric when reflecting their trajectories about an appropriate axis leaves
their statistical behavior unchanged. Because reflection about any axis through
$\bm c$ reverses the sign of $j$ and the direction of rotation, rotation can be quantified by a measure of
dynamical mirror-symmetry breaking (Fig.~\ref{fig:rca-overview}A and Extended Data
Fig.~\ref{fig:extended-symmetry}A). We refer to this quantity as the asymmetry
production rate (APR); the formal definition extends to other spatial transformations
(see Methods). In practice, we estimate APR from an ensemble of trajectory segments
$\Gamma$, each spanning two adjacent samples separated by a time interval $\tau$. The
accumulated angular current along each segment is
$J_\tau[\Gamma]=\int_0^\tau j(t)\,\mathrm{d}t$, and the estimator is
\begin{equation}
\widehat\sigma(\tau)=\frac{1}{\tau}
\frac{2\overline{J_\tau}^{\,2}}{\Delta J_\tau},
\label{eq:apr-estimator}
\end{equation}
where $\overline{J_\tau}$ and $\Delta J_\tau$ are the mean and variance, respectively,
across the ensemble of segments. Persistent directed
rotation produces a nonzero mean current, whereas rotation-irrelevant fluctuations
increase its variance; $\widehat\sigma$
therefore balances persistent directional current against rotation-irrelevant
fluctuations.

We constructed rotational component analysis (RCA) by searching over pairs of
orthonormal projection vectors $\bm w_1$ and $\bm w_2$ to maximize
$\widehat\sigma$. The optimization was performed using gradient descent and automatic
differentiation. RCA
therefore identifies the plane that maximizes this current-to-fluctuation measure (Fig.~\ref{fig:rca-overview}A). This
objective differs from that of principal component analysis
(PCA), which selects directions of maximal variance. A rotational mode can
be missed by PCA when it occupies a low-variance plane approximately orthogonal to the leading
principal component (Fig.~\ref{fig:rca-overview}B). RCA is unsupervised by construction
because no task variables or behavioral labels enter the optimization.

We benchmarked RCA using motor-cortical spiking activity during reaching, a setting in
which rotational population dynamics are well established~\cite{churchland2012neural}.
Previous studies typically extracted these rotations from activity averaged across
repeated trials within each reach condition, partly due to the reduced rotation-irrelevant fluctuations in the averaged trajectories~\cite{churchland2012neural}. We first
reproduced this conventional analysis by fitting RCA and jPCA to condition-averaged
activity. jPCA is a PCA-based method that first projects population activity onto a leading principal-component subspace and then extracts rotational dynamics by fitting a skew-symmetric linear dynamical system~\cite{churchland2012neural}.  Both methods recovered rotational trajectories in the macaque and human
datasets (Fig.~\ref{fig:rca-overview}C,D,I,J).

To determine whether these rotational modes remained detectable in the presence of
trial-to-trial variability, we next fitted RCA and jPCA directly to individual-trial
trajectories pooled across reach conditions, without averaging trials before
dimensionality reduction (see Methods). For visualization, we projected each trial into
the fitted plane and subsequently averaged the projected trajectories within each
condition. In the macaque data, the RCA plane fitted to individual-trial activity retained
coherent loops across reach conditions, whereas the corresponding jPCA plane did not
show coherent rotational organization (Fig.~\ref{fig:rca-overview}E,F). In the human
data, both planes fitted to individual-trial activity contained visible rotational
trajectories (Fig.~\ref{fig:rca-overview}K,L), motivating a direct comparison of their
consistency across individual trials.

We quantified rotation on each trial using the normalized cumulative angular
current
\begin{equation}
\begin{aligned}
\mathcal{J}_i(t)
&=\frac{1}{R^2} \int_{t_0}^{t}\!j_i(s;\bm c)\,\diff s,
\end{aligned}
\label{eq:normalized-cumulative-current}
\end{equation}
where $i$ indexes trials, $j_i(s;\bm c)$ is the angular current at time $s$, $\bm c$ is the empirical center, and $R^2$ is the mean-squared
radius across trials and time. 
We defined the trial-to-trial signal-to-noise ratio (SNR) at the final time $t_f$ as
\begin{equation}
\mathrm{SNR}
=\frac{\left|\left\langle\mathcal{J}_i(t_f)\right\rangle_i\right|}
{\operatorname{SD}_i\!\left[\mathcal{J}_i(t_f)\right]},
\label{eq:trial-current-snr}
\end{equation}
where $\operatorname{SD}_i$ denotes the sample standard deviation across trials. The
SNR measures the consistency of rotation across individual
trials.
For macaque reaching behavior, RCA produced a coherent cumulative current (SNR, 0.89),
whereas the jPCA current remained near zero (SNR, 0.05;
Fig.~\ref{fig:rca-overview}G,H). For  human reaching behavior,
the current increased in both projections, but was more consistent across trials in the
RCA plane (SNR, 1.90) than in the jPCA plane (SNR, 0.98;
Fig.~\ref{fig:rca-overview}M,N). Thus, RCA better recovered rotational activity directly from
non-averaged population activity and identified a shared rotational coordinate that was
consistently expressed across individual trials. We show in the Methods that RCA can also identify rotational modes in non-neural datasets (Extended Data Fig.~\ref{fig:extended-cycle}).

\subsection*{A brain-wide rotational mode predicts mouse responses to a random cue}

\begin{figure*}[!t]
\centering
\includegraphics[width=\textwidth]{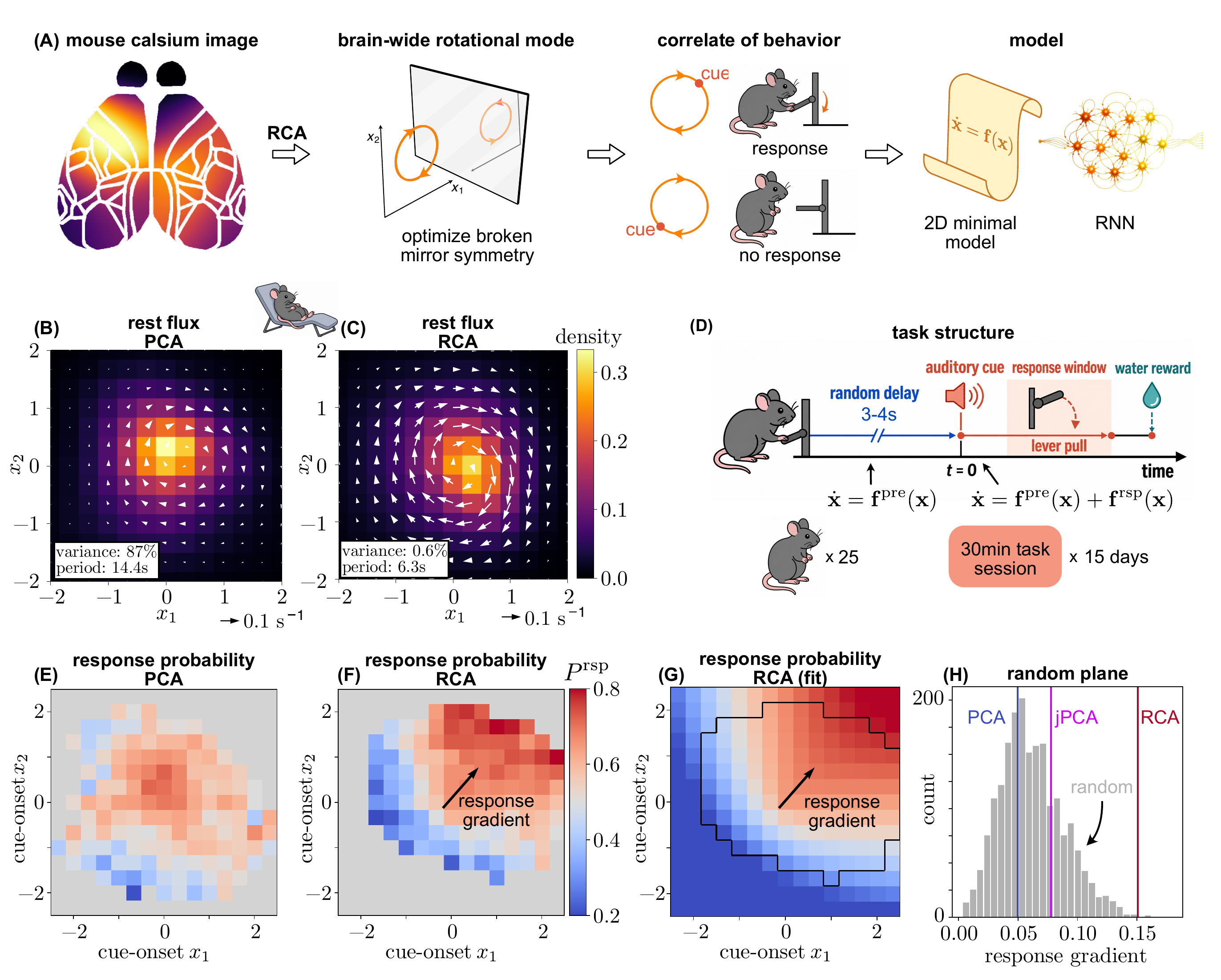}
\caption{\textbf{Brain-wide rotational mode predicts cue-evoked behavior in mice.}
\textbf{A}, Analysis overview. Cortex-wide calcium activity was projected by rotational
component analysis (RCA), which optimizes broken mirror symmetry, onto a plane with a
circulating probability current. The state at cue onset distinguished response from
nonresponse trials. This behavioral association was reproduced by a two-dimensional
minimal model and a recurrent neural network (RNN). \textbf{B,C}, Resting-state probability density (color)
and probability flux (white arrows) in the PCA (\textbf{B}) and RCA (\textbf{C})
planes. Annotations report the fraction of total variance explained by each
projection and the estimated rotation period. Translucent white ellipses indicate the standard errors of the
mean for the flux vectors; most are smaller than the arrowheads at this plotting scale.
\textbf{D}, Lever-pull task. Mice
were trained to respond to an auditory cue; a pull during the response window was
followed by a water reward. The cue appeared after a random delay of 3--4 s following
the preceding lever pull. The dataset comprised 25 mice recorded in 30-min task
sessions over 15 training days. \textbf{E,F}, Empirical response probability
conditioned on cue-onset state in the PCA (\textbf{E}) and RCA (\textbf{F}) planes.
The arrow in \textbf{F} shows the fitted response-gradient direction. \textbf{G},
Response-probability surface fitted in the RCA plane; the arrow again indicates the
directional gradient. \textbf{H}, Distribution of response-gradient magnitudes across
2,000 random planes. Vertical lines indicate the PCA ($g=0.050$), jPCA ($g=0.078$)
and RCA ($g=0.151$) planes. All projection planes were orthonormalized and their
projected coordinates were scaled to unit variance before fitting (Methods).}
\label{fig:mouse}
\end{figure*}

We next applied APR to wide-field calcium imaging of mouse brain activity. We analyzed
the BraiDyn-BC dataset, which contains longitudinal recordings from 25 head-fixed mice
expressing GCaMP6s in cortical excitatory neurons during rest and during acquisition of
a tone-triggered lever-pull task over 15 training days~\cite{kondo2025multimodal}. We used
hemodynamics-corrected fluorescence sampled at 30 Hz from 44 regions (22 per hemisphere) registered to the
Allen Common Coordinate Framework (Fig.~\ref{fig:mouse}A, left). The regions span the mouse dorsal cortex, including motor, somatosensory, olfactory, visual, auditory, and retrosplenial areas (Extended Data Fig.~\ref{fig:extended-mode-analysis}).

We first compared resting-state probability flux in the PCA and RCA planes
(Fig.~\ref{fig:mouse}B,C). In the PCA plane, where $x_1$ and $x_2$ are the first two
principal components, the probability flux showed only weak circulation
(Fig.~\ref{fig:mouse}B). By contrast, the RCA plane revealed a strong clockwise
probability current (Fig.~\ref{fig:mouse}C). This circulation was identified from
spontaneous activity and therefore did not depend on sensory stimulation or task
events. Notably, the RCA plane accounted for only $0.6\%$ of the total variance,
compared with $87\%$ for the PCA plane, highlighting how a rotational mode can
be overlooked by conventional dimensionality-reduction methods. The regional
composition of this mode and the minimal antisymmetric coupling pattern show
loadings across the cortex, with large values in motor and somatosensory areas (Extended Data Fig.~\ref{fig:extended-mode-analysis}). This latent
rotation coordinates distributed brain regions that need not be spatially adjacent,
distinguishing it from topographic rotating waves~\cite{ye2026rotating}.

We next asked whether the rotational mode identified at rest was related to behavior
during task sessions. An auditory cue initiated a response window in which a
successful lever pull was followed by a water reward (Fig.~\ref{fig:mouse}D). Each cue
occurred after a random delay of 3--4 s following the preceding lever pull. This
randomized timing sampled different states of the ongoing cortical
dynamics across trials, allowing us to test whether the cue-onset state predicted task
outcome. We defined $P^{\rm rsp}(\bm x^{\rm cue})$ as the probability that the mouse
responds to a cue, conditioned on its state $\bm x^{\rm cue}$ in the reduced plane at
cue onset. Response
probability varied only weakly and irregularly across the PCA plane
(Fig.~\ref{fig:mouse}E), but showed a strong directional dependence in the RCA plane:
cues that occurred when cortical activity occupied the upper-right portion of the plane
were followed by responses more often than cues that occurred when activity occupied
the lower-left portion
(Fig.~\ref{fig:mouse}F). We described this pattern as a radially symmetric baseline
plus a directional linear component,
\begin{equation}
P^{\mathrm{rsp}}(\bm x^{\rm cue})=P_0(1-c_0\lVert\bm x^{\rm cue}\rVert)+\bm g\!\cdot\!\bm x^{\rm cue}, 
\label{eq:directional-gradient}
\end{equation}
where $\bm x^{\rm cue}$ is the cue-onset state and $P_0$, $c_0$ and $\bm g$ were fitted
to the empirical data. The fit captured both
the directional increase and the weaker radial dependence of response probability
(Fig.~\ref{fig:mouse}G). Here, $\bm g$ is the response gradient and points in the
direction of increasing response probability.

To determine whether this predictive relationship was unusual among two-dimensional
projections, we compared the magnitude of the response gradient with those obtained
from 2,000 random planes (Fig.~\ref{fig:mouse}H). The RCA gradient lay near the extreme
upper tail of the random-plane distribution and was substantially larger than the PCA
and jPCA gradients. Notably, the RCA plane was selected using only resting-state
activity, without access to task or behavioral information. Thus, a spontaneous
rotational mode identified without behavioral supervision nevertheless provided a
strong neural correlate of subsequent cue-evoked behavior. We further asked whether
this predictive relationship could be explained by arousal, for which pupil diameter
provides a widely used proxy. On held-out sessions, the cue-onset RCA state predicted
trial outcome better than pupil-derived arousal (Extended Data
Fig.~\ref{fig:extended-pupil}).

\subsection*{Minimal and recurrent-network models explain state-dependent responses}

To identify a heuristic dynamical mechanism linking cue-onset state to behavior, we first
constructed a two-dimensional minimal model of brain activity
(Fig.~\ref{fig:model}A):
\begin{equation}
\diff \bm x_t = \bm f(\bm x_t,t)\,\diff t + \diff \bm W_t,
\label{eq:drift}
\end{equation}
where $\bm f(\bm x_t,t)$ is the state- and time-dependent drift and $\bm W_t$ is a
two-dimensional Wiener process representing stochastic fluctuations. Cue onset changes
the drift field: before the cue, activity follows the pre-cue (i.e., intrinsic) drift
$\bm f^{\mathrm{pre}}$, whereas after the cue it follows the post-cue drift
$\bm f^{\mathrm{post}}$. We decomposed the latter into the intrinsic drift and a
cue-evoked response component,
\begin{equation}
\bm f^{\mathrm{post}}(\bm x)=\bm f^{\mathrm{pre}}(\bm x)
+\bm f^{\mathrm{rsp}}(\bm x).
\label{eq:drift-sum}
\end{equation}
We estimated the empirical pre-cue and post-cue drift fields over windows of ten and five
frames, respectively. The pre-cue field preserved the clockwise rotation identified at
rest (Fig.~\ref{fig:model}B). By contrast, the cue-evoked response drift pointed
approximately along a common direction $\hat{\bm n}_{\parallel}$, whereas its magnitude
varied primarily along the orthogonal direction $\hat{\bm n}_{\perp}$
(Fig.~\ref{fig:model}C). We captured this organization with the two-dimensional vector field:
\begin{figure*}[!t]
\centering
\includegraphics[width=\textwidth]{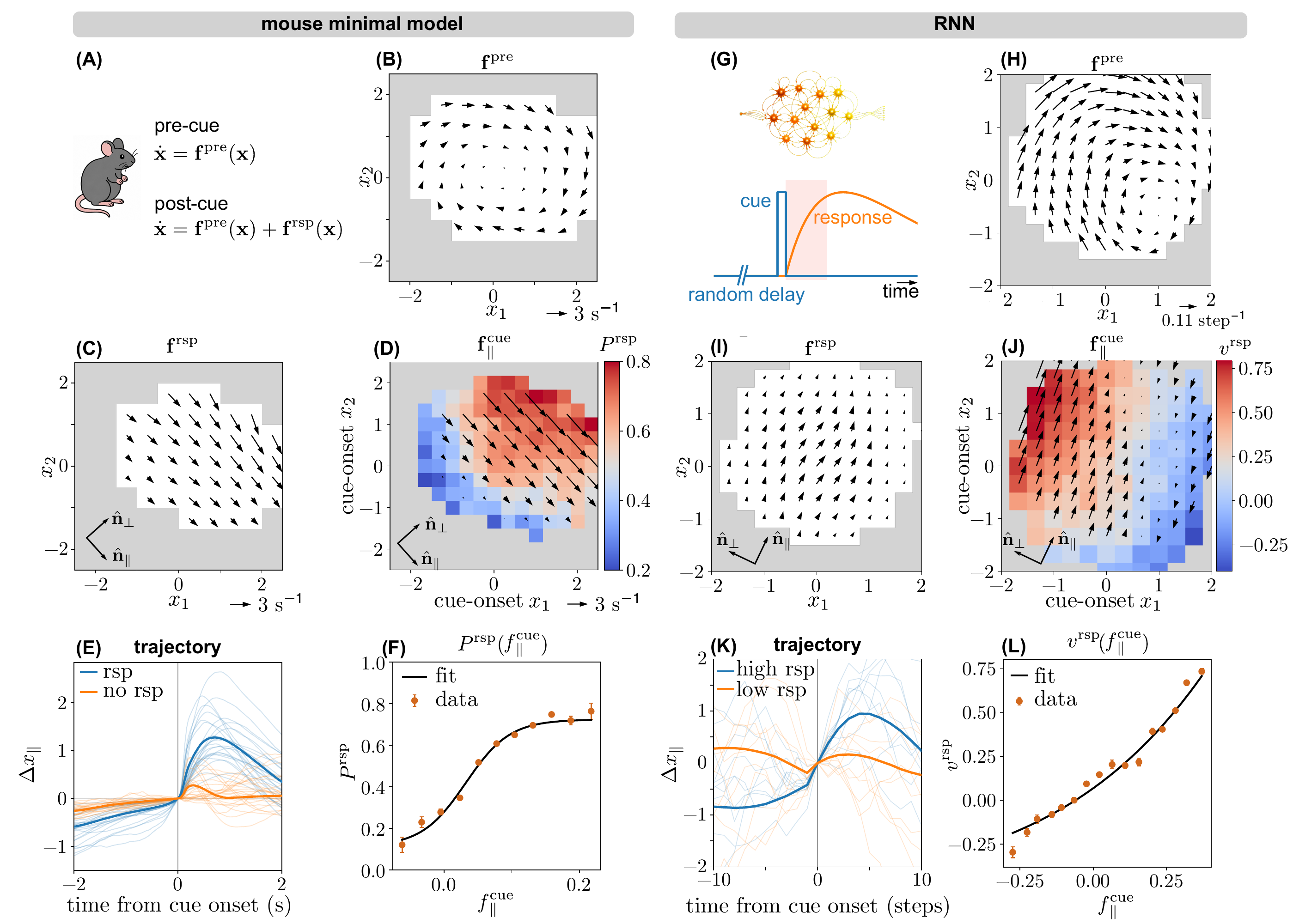}
\caption{\textbf{Minimal dynamical model and an RNN reproduce the state-dependent
response rule.} \textbf{A}, Mouse minimal model. Pre-cue activity follows
the intrinsic rotational drift $\bm f^{\mathrm{pre}}$; adding the cue-evoked response
drift $\bm f^{\mathrm{rsp}}$ gives the post-cue drift $\bm f^{\mathrm{post}}$.
\textbf{B}, Empirical
pre-cue drift field $\bm f^{\mathrm{pre}}$ in the rotational plane. \textbf{C},
Empirical cue-evoked response-drift field $\bm f^{\mathrm{rsp}}$. Basis arrows define the response direction
$\hat{\bm n}_{\parallel}$ and the orthogonal direction $\hat{\bm n}_{\perp}$.
\textbf{D}, Response probability $P^{\mathrm{rsp}}$ (color)
conditioned on the cue-onset state, overlaid with the post-cue drift
component, $\bm f^{\mathrm{post}}_{\parallel} =f^{\mathrm{post}}_{\parallel}\hat{\bm n}_{\parallel}$
(black arrows).
\textbf{E}, Trial-averaged trajectories (dark curves) projected along
$\hat{\bm n}_{\parallel}$ and aligned to cue onset, separated by response outcome;
translucent curves denote trajectories of each individual mouse. \textbf{F}, Response probability as a function of
the post-cue drift component $f^{\mathrm{post}}_{\parallel}$. Points show binned data
and the black curve shows the fitted response rule. \textbf{G}, Recurrent neural
network (RNN) trained to produce a graded response following a randomly timed cue. The
shaded region indicates the response window.
\textbf{H}, Pre-cue rotational drift field generated by the RNN. \textbf{I},
Cue-evoked response-drift field generated by the RNN. \textbf{J}, RNN response
strength $v^{\mathrm{rsp}}$ (color) conditioned on the cue-onset state, overlaid with
$f^{\mathrm{post}}_{\parallel}\hat{\bm n}_{\parallel}$ (black arrows).
\textbf{K}, Trial-averaged RNN trajectories (dark curves) projected along
$\hat{\bm n}_{\parallel}$ and aligned to cue onset, separated into high- and
low-response trials; translucent curves denote 10 example trials. \textbf{L}, Response strength
as a function of $f^{\mathrm{post}}_{\parallel}$. Points show binned simulations and
the black curve shows the fit. In \textbf{B,C,H,I}, arrows indicate the direction and
magnitude of the corresponding drift fields.}
\label{fig:model}
\end{figure*}
\begin{equation}
\bm f^{\mathrm{rsp}}(\bm x)=f_0^{\rm rsp}
\left(1+c\,\bm x\!\cdot\!\hat{\bm n}_{\perp}\right)
\hat{\bm n}_{\parallel}.
\label{eq:response-drift}
\end{equation}

To relate this drift geometry to behavior, we evaluated the post-cue drift at each
cue-onset state and projected it onto the response direction,
$f^{\mathrm{post}}_{\parallel}=\bm f^{\mathrm{post}}(\bm x^{\mathrm{cue}})\cdot
\hat{\bm n}_{\parallel}$. The corresponding vector component is
$\bm f^{\mathrm{post}}_{\parallel}=f^{\mathrm{post}}_{\parallel}
\hat{\bm n}_{\parallel}$. The scalar $f^{\mathrm{post}}_{\parallel}$ increased
predominantly along the transverse direction $\hat{\bm n}_{\perp}$
(Fig.~\ref{fig:model}D). This transverse gradient is expected from the intrinsic
rotational flow: its response-direction component,
$f^{\mathrm{pre}}_{\parallel}=\bm f^{\mathrm{pre}}\cdot
\hat{\bm n}_{\parallel}$, is positive in one half of the plane and negative in the
other. Notably, the behavioral gradient $\bm g$ was also approximately aligned with
$\hat{\bm n}_{\perp}$, and response probability was higher at states with larger
$f^{\mathrm{post}}_{\parallel}$ (Fig.~\ref{fig:model}D). Thus, the response can be predicted from $f^{\mathrm{post}}_{\parallel}$, and cues arriving at
different phases of the rotation encounter different response probabilities.

We tested this prediction using cue-aligned trajectories from response and nonresponse
trials (Fig.~\ref{fig:model}E). Both groups showed an abrupt increase in displacement
along $\hat{\bm n}_{\parallel}$ during the post-cue period, but the increase was substantially
larger on response trials. Consistent with the model, empirical response probability
increased monotonically from below 20\% to above 70\% across the observed range of
$f^{\mathrm{post}}_{\parallel}$ (Fig.~\ref{fig:model}F). This dependence was well
described by a sigmoid:
\begin{equation}
\begin{aligned}
P^{\rm rsp}(f^{\mathrm{post}}_{\parallel})={}&P_{\text{low}}
+\left(P_{\text{high}}-P_{\text{low}}\right)\\
&\times\left[1+e^{-\lambda\left(f^{\mathrm{post}}_{\parallel}-s_0\right)}\right]^{-1}.
\end{aligned}
\label{Prsp}
\end{equation}
Thus, the interaction between the rotational drift and the cue-evoked drift provides a
compact account of the state-dependent response probability.

We next asked whether a similar relation between response and rotation could also emerge in a simpler artificial system: a recurrent dynamical
system trained to generate a randomly cued response. We used a recurrent neural network
(RNN) with dynamics
\begin{equation}
\begin{aligned}
\bm h_{t+1}={}&(1-\alpha)\,\bm h_t
+\alpha\,\tanh\!\left(\bm W_{\mathrm{rec}}\bm h_t
+\bm W_{\mathrm{in}}u_t+\bm b\right)\\
&+\sqrt{2\alpha}\,\sigma\,\bm \xi_t,
\end{aligned}
\label{eq:rnn-dynamics}
\end{equation}
where $\bm h_t$ is the network state, $\bm W_{\mathrm{rec}}$ is the recurrent weight
matrix, $\bm W_{\mathrm{in}}$ maps the scalar cue input $u_t$ into the network,
$\bm b$ is the bias vector, $\alpha$ is the leak parameter, and $\sigma$ is the
noise amplitude with $\bm \xi_t\sim\mathcal{N}(0,I)$. The model reproduced the essential
temporal structure of the mouse task~\cite{kondo2025multimodal} (Fig.~\ref{fig:model}G; see Methods): a pulse in
$u_t$ at a random time represented the randomly timed auditory cue, and the fixed input
weights $\bm W_{\mathrm{in}}$ specified how that cue entered the network. The hidden
state $\bm h_t$ represented population activity, whose internal dynamics were governed
by $\bm W_{\mathrm{rec}}$ and $\bm b$. Activity was read out along a fixed direction
$\bm W_{\mathrm{out}}$, and response strength was defined by the normalized net output
displacement from the beginning to the end of the fixed post-cue response window. Thus,
$\bm W_{\mathrm{in}}$ and
$\bm W_{\mathrm{out}}$ provided fixed input and output mappings, whereas training by
backpropagation modified only $\bm W_{\mathrm{rec}}$ and $\bm b$ to increase response
strength.

After training, we applied RCA to the spontaneous network dynamics in the absence of a
cue and estimated the intrinsic and cue-evoked drift fields as in the mouse data. The
RCA plane showed a pre-cue rotation and a cue-evoked drift oriented approximately along
a common direction $\hat{\bm n}_{\parallel}$ (Fig.~\ref{fig:model}H,I). Response
strength varied directionally across cue-onset states. As in the mouse data, overlaying
$\bm f^{\mathrm{post}}_{\parallel}$ on the response-strength map showed that larger
responses occurred in states with larger $f^{\mathrm{post}}_{\parallel}$; the response-strength gradient was approximately
perpendicular to $\hat{\bm n}_{\parallel}$ (Fig.~\ref{fig:model}J). Moreover, high- and low-response
trials followed distinct trajectories along $\hat{\bm n}_{\parallel}$
(Fig.~\ref{fig:model}K), and response strength increased with
$f^{\mathrm{post}}_{\parallel}$ (Fig.~\ref{fig:model}L), qualitatively reproducing the
relationships observed in mice (Fig.~\ref{fig:model}E,F). Thus, a recurrent system with
no explicitly imposed rotational structure can reproduce the key geometry linking
spontaneous flow and cue-onset state to response strength, as observed in mice.

\subsection*{Learning strengthens the cue-evoked response drift}
\begin{figure*}[!t]
\centering
\includegraphics[width=\textwidth]{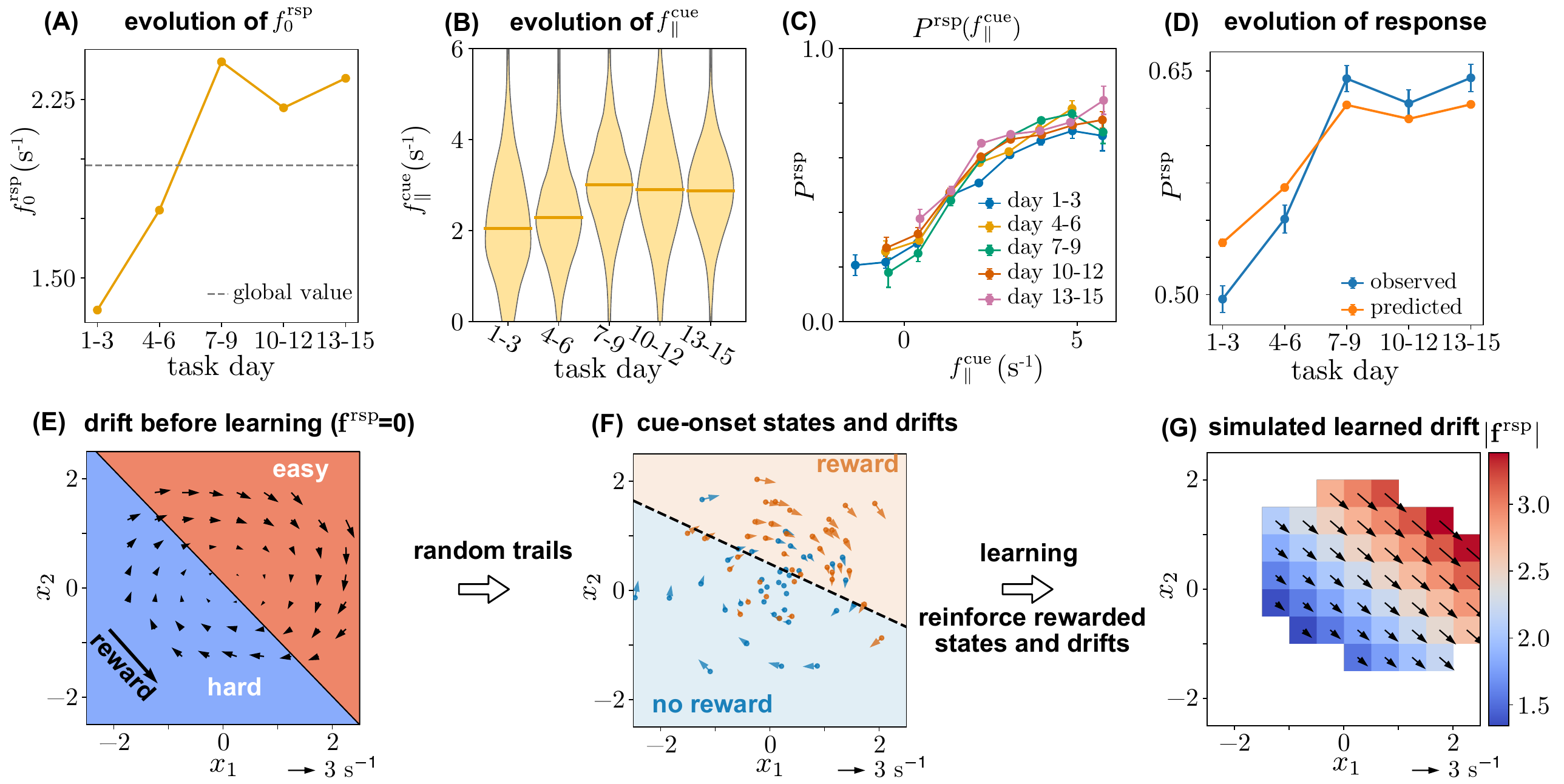}
\caption{\textbf{Learning strengthens the cue-evoked response drift within an existing
rotational coordinate.} \textbf{A--D}, Learning across 15 task days. \textbf{A},
Inferred response-drift amplitude $f^{\mathrm{rsp}}_0$ in successive three-day windows.
The dashed line indicates the estimate from all task days. \textbf{B}, Distribution of
$f^{\mathrm{post}}_{\parallel}$ in each window; horizontal bars indicate medians.
\textbf{C}, Empirical response probability as a function of
$f^{\mathrm{post}}_{\parallel}$ for each training window. Symbols and error bars show
binned estimates and their uncertainty. \textbf{D}, Observed response probability and
the cross-validated probability predicted from window-specific drift estimates and a
fixed response rule. Symbols and error bars show estimates and their uncertainty.
\textbf{E--G}, Proposed learning mechanism. \textbf{E}, Intrinsic drift before learning
($\bm f^{\mathrm{rsp}}=0$). The diagonal boundary separates states with relatively high
(``easy'') and low (``hard'') response probability; the arrow indicates the rewarded
direction. \textbf{F}, Cue-onset states and drifts from randomly sampled trials,
colored by reward outcome. The dashed line is the optimal linear boundary separating
rewarded and unrewarded trials. \textbf{G}, Simulated response drift generated by
reinforcing rewarded states and drifts; arrows indicate direction and color denotes
$|\bm f^{\mathrm{rsp}}|$.}
\label{fig:learning}
\end{figure*}

The lever-pull response is a learned rather than an innate mouse behavior. We therefore asked how the
response drift changed as the behavior was acquired. We inferred
$\bm f^{\mathrm{pre}}$ and $\bm f^{\mathrm{rsp}}$ in five nonoverlapping three-day
windows spanning the 15 training days. The intrinsic drift $\bm f^{\mathrm{pre}}$
was largely unchanged, whereas the cue-evoked
field $\bm f^{\mathrm{rsp}}$ showed a general increase
in magnitude (Extended Data Fig.~\ref{fig:extended-learning}). This is evident in the amplitude parameter $f^{\mathrm{rsp}}_0$, which shows an overall increasing trend across training (Fig.~\ref{fig:learning}A). In parallel, the
distribution of $f^{\mathrm{post}}_{\parallel}$ shifted toward the response-generating
regime during learning (Fig.~\ref{fig:learning}B), whereas the conditional drift-to-response mapping
$P^{\rm rsp}(f^{\mathrm{post}}_{\parallel})$ remained similar across training windows
(Fig.~\ref{fig:learning}C). A cross-validated model that held this response rule fixed
while allowing the inferred response drift to vary captured most of the improvement in
observed performance (Fig.~\ref{fig:learning}D). These results suggest that learning
primarily alters the cue-evoked dynamics, and hence the distribution of
$f^{\mathrm{post}}_{\parallel}$, while preserving the mapping from this dynamical
variable to response probability.

During learning, the direction of the transverse gradient of $\bm f^{\rm rsp}$ remained fixed, which is consistent with the intrinsic rotational dynamics. This gradient was not captured by the RNN (Fig.~\ref{fig:model}I). This difference may reflect the distinct learning signals available
to the two systems. The RNN was trained by backpropagation using an explicit,
differentiable objective, whereas the mice presumably received sparse trial-level reward feedback. 
To explain how sparse reward feedback could produce this, we constructed a
minimal reinforcement model. 

Before a learned response drift develops, the intrinsic
rotation alone occasionally carries activity through states whose velocity points along
the rewarded response direction (Fig.~\ref{fig:learning}E). Because cues occur at
random times, they sample both favorable and unfavorable phases of the rotation,
producing rewarded and unrewarded trials. The states and drifts associated with reward
therefore form a biased subset of all samples in the rotational plane
(Fig.~\ref{fig:learning}F). Reinforcing this subset could progressively shape a
state-dependent cue-evoked response field. We represented that field in the minimal form
\begin{equation}
\bm f^{\mathrm{rsp}}(\bm x)=\bm f^{\rm rsp}_0(1+\bm k\cdot\bm x),
\label{eq:learning-field}
\end{equation}
where $\bm f^{\rm rsp}_0$ sets the mean response drift, $f^{\mathrm{rsp}}_0=\lVert\bm f^{\rm rsp}_0\rVert$
denotes its amplitude and $\bm k$ determines its state dependence. For each rewarded
sample with state $\bm x^s$ and drift $\bm f^s$, we updated these parameters according to
\begin{equation}
\bm f^{\rm rsp}_0\leftarrow(1-a_1)\bm f^{\rm rsp}_0+b_1\bm f^s,\qquad
\bm k\leftarrow(1-a_2)\bm k+b_2\bm x^s.
\label{eq:learning-rule}
\end{equation}
The decay terms stabilized the parameters, whereas the reward-gated terms reinforced
the drifts and locations associated with successful trials. Simulating this rule with
the inferred pre-cue dynamics produced an anisotropic response field resembling that
observed in mice (compare Fig.~\ref{fig:learning}G to Fig.~\ref{fig:model}C; see Methods). In this model, learning does
not create a new dynamical coordinate. Instead, it strengthens a cue-evoked drive within
the coordinate supplied by spontaneous rotation, while the intrinsic flow constrains
which states and drifts are reinforced. The model thus provides a possible account of
how animals can acquire the response without access to an explicit cost function.

\subsection*{Rotational currents generalize to humans and detect early psychosis}

\begin{figure}[!t]
\centering
\includegraphics[width=0.5\textwidth]{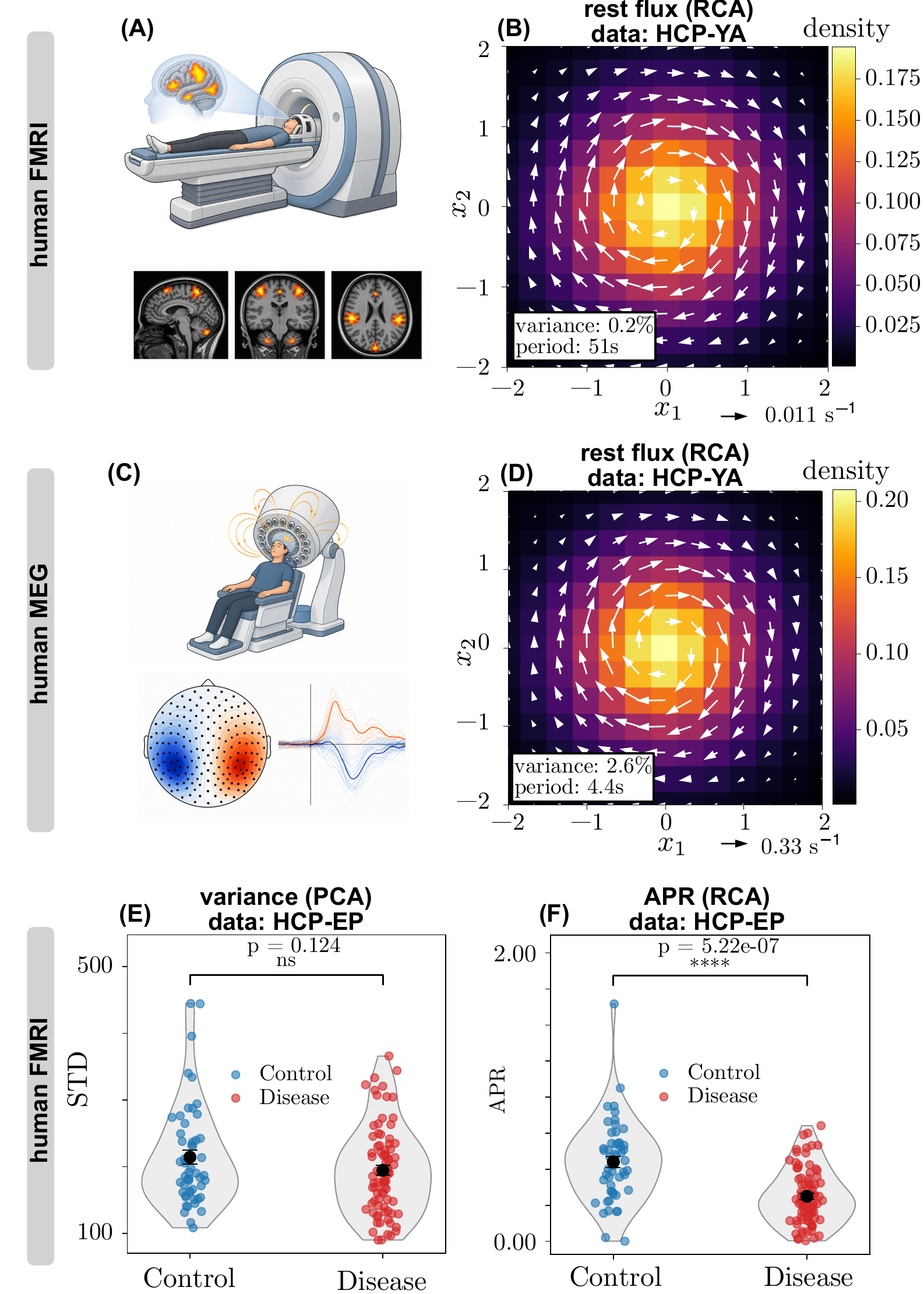}
\caption{\textbf{Brain-wide rotational dynamics differ in early psychosis.}
\textbf{A}, Schematic of human fMRI acquisition and illustrative orthogonal activation
maps. \textbf{B}, Resting-state probability density (color) and probability flux
(white arrows) in the RCA plane for the HCP Young Adult (HCP-YA) fMRI dataset.
The annotation reports the fraction of total variance explained by the RCA plane and
the estimated rotation period.
\textbf{C}, Schematic of human MEG acquisition with an illustrative sensor
topography and multichannel evoked responses. \textbf{D}, Resting-state probability
density and probability flux in the RCA plane for the HCP-YA MEG dataset. The annotation
reports the fraction of total variance explained and the estimated rotation period.
\textbf{E,F}, Participant-level PCA variability, defined as the standard deviation along the leading principal axis (\textbf{E}), and RCA asymmetry production
rate (APR; \textbf{F}) in control participants and participants with early psychosis
from the HCP Early Psychosis (HCP-EP) dataset.
Points indicate participants, violins show the group distributions, and black symbols
and bars show group summaries. Brackets indicate group-comparison $P$ values.
Scale arrows in \textbf{B} and \textbf{D} give the
vector scale.}
\label{fig:human}
\end{figure}

We next tested whether rotational modes can be generalized beyond mouse cortex and calcium imaging. We
analyzed resting-state recordings from a large cohort of healthy young adults in the
Human Connectome Project Young Adult (HCP-YA) dataset~\cite{vanessen2013wuminn}, which includes whole-brain fMRI
and MEG data. Applying RCA to the HCP-YA fMRI recordings revealed a rotational current in a
projection of cortical activity (Fig.~\ref{fig:human}A,B). HCP-YA MEG recordings
likewise contained a coherent rotational current (Fig.~\ref{fig:human}C,D), despite MEG's
distinct spatial resolution and much faster sampling rate. The recurrence of organized
currents in mouse calcium imaging, human fMRI and human MEG indicates that
symmetry-guided projection is not restricted to one species, recording modality or
timescale. The inferred rotation periods differed markedly, from 51 s in fMRI to 4.4 s
in MEG, consistent with the distinct temporal resolutions and characteristic signal
dynamics of the two modalities. Despite this difference, the rotational planes
accounted for only 0.2\% of the variance in fMRI and 2.6\% in MEG, in agreement with the
low-variance rotational mode identified in mouse cortex.

We then asked whether rotational irreversibility could act as a probe for psychiatric disorders.
For this comparison, we used resting-state fMRI from the HCP Early Psychosis
(HCP-EP) dataset, which includes control participants and participants with early
psychosis~\cite{hancock2023metastability}. Variability along the leading principal component did not differ significantly
between groups ($P=0.124$; Fig.~\ref{fig:human}E). By contrast, the APR of the RCA projection was markedly lower in the early-psychosis group
($P=5.22\times10^{-7}$; Fig.~\ref{fig:human}F). Participant-level split-sample
analyses confirmed that reduced APR was robust to the group used for projection
learning, whereas PCA variability showed at most a modest group difference (Extended
Data Fig.~\ref{fig:extended-hcpep}; see Methods). The rotational measure therefore revealed a group difference that was not
apparent from PCA variability alone: reduced brain-wide rotational dynamics were
associated with early psychosis.

\subsection*{Double-rotation modes separate preparation and predict future responses}
We further generalized RCA to detect more complex dynamical modes. Here, we
targeted double-rotational dynamics, characterized by oppositely directed circulation
around neighboring centers, using double-rotational component analysis (DRCA).
An ideal pair of counter-rotating currents can be mirror-symmetric. By contrast, the pair breaks half-turn symmetry: a $180^\circ$ rotation
exchanges the two cycles but preserves chirality and therefore does not reproduce the
original current pattern (Extended Data Fig.~\ref{fig:extended-symmetry}B). Consistent with this distinction, under a half-turn about the
empirical center, the ordinary angular current $j$ is unchanged, whereas the spatially
weighted current $j^*=(x_1-c_1)j$ changes sign. We therefore detect double-rotation modes by optimizing
\begin{equation}
\widehat\sigma_{\mathrm{double}}=\frac{1}{\tau}\frac{2\overline{J_\tau^*}^{\,2}}{\Delta J_\tau^*},
\label{eq:double}
\end{equation}
where $\bm c=[c_1,c_2]$ is the empirical center.
The factor $x_1-c_1$ changes sign across the midline, causing counter-rotating currents
on the two sides to contribute with the same sign rather than cancel in the ordinary
angular current.

\begin{figure}[!t]
\centering
\includegraphics[width=0.5\textwidth]{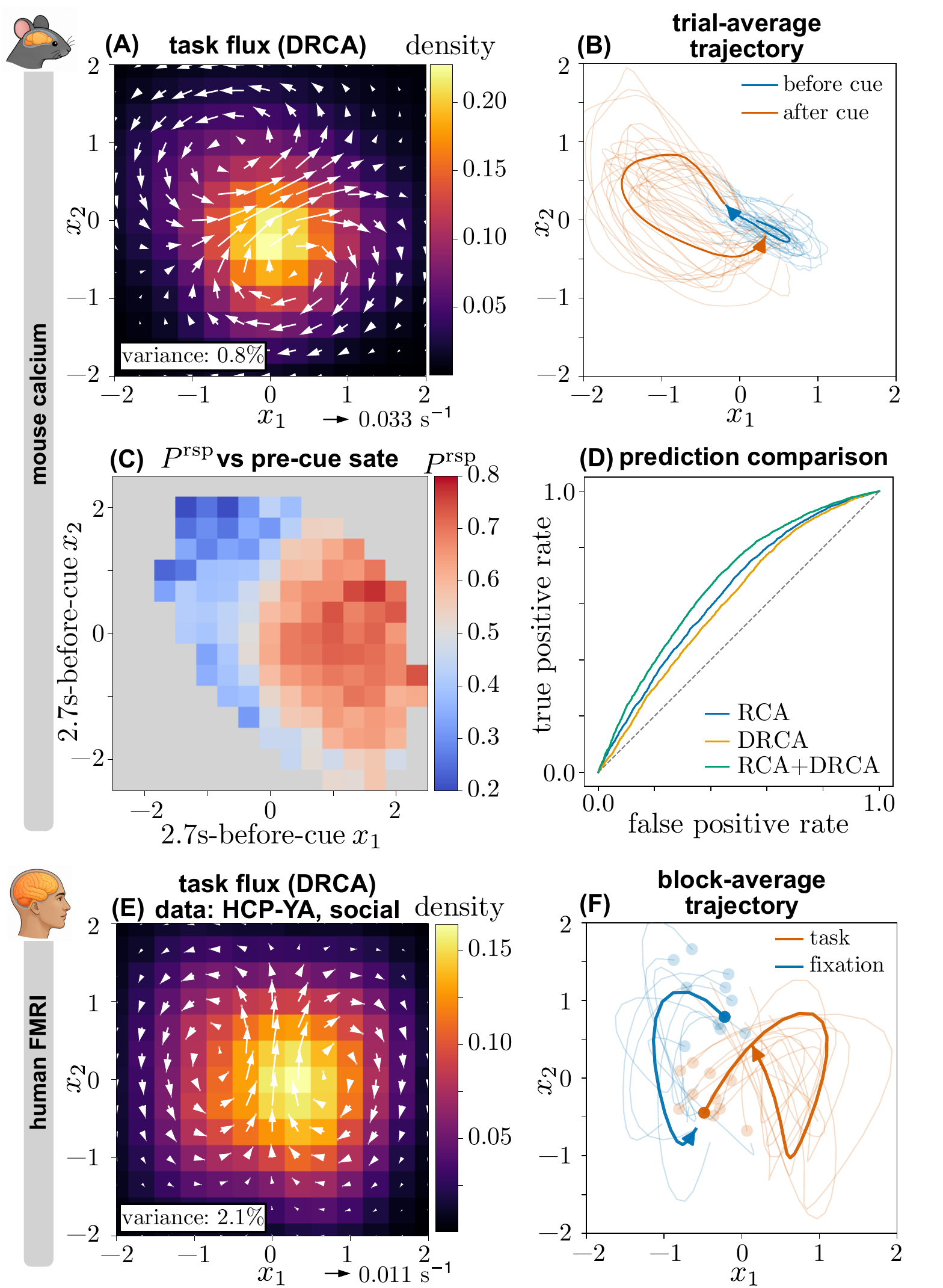}
\caption{\textbf{Double-rotation modes separate preparatory and task-related
dynamics.} \textbf{A}, Mouse task probability density (color) and probability flux
(white arrows) in the double-rotational component analysis (DRCA) plane. The annotation
reports the fraction of total variance explained. 
\textbf{B}, Pre-cue (blue) and post-cue (orange) mouse trajectories aligned to cue onset.
Translucent curves show trial-averaged trajectories for individual mice, thick curves show group averages,
and arrowheads indicate temporal direction. \textbf{C}, Mouse response probability
conditioned on the DRCA state 2.7 s pre-cue; gray bins were excluded because
of insufficient occupancy. \textbf{D}, Receiver operating characteristic curves for
classifiers evaluated on held-out sessions using the cue-onset RCA state
(AUC $=0.638$), pre-cue DRCA state (AUC $=0.612$), or their combination
(AUC $=0.677$). State--response mappings and classifiers were fitted using separate
training sessions. The diagonal denotes chance performance. \textbf{E},
Probability density and probability flux for the DRCA mode identified in the HCP-YA
human fMRI social task. The annotation
reports the fraction of total variance explained. \textbf{F}, Block-aligned trajectories during task (orange) and
fixation (blue). Light curves show block-averaged trajectories for 10 randomly selected
participants, thick curves show group averages, circles mark aligned time points, and arrowheads indicate temporal
direction.}
\label{fig:double}
\end{figure}

Applied to mouse task activity, DRCA revealed a probability flux with two distinct
centers of circulation (Fig.~\ref{fig:double}A). Cue-aligned trajectories showed that
the two centers organized different periods of each trial: during the pre-cue period, activity
circulated around the lower-right center, whereas during the post-cue period it followed a broader
loop around the upper-left center (Fig.~\ref{fig:double}B). The double-rotation mode
therefore separated ongoing preparatory activity from post-cue and response-related
dynamics.

The preparatory branch of the double rotation carried information about the animal's
future response. Response probability depended strongly on the projected cortical state
80 frames, or 2.7 s, pre-cue (Fig.~\ref{fig:double}C). Because cue timing was
randomized, this dependence cannot be attributed to a deterministic countdown to a
predictable cue time. Using separate training and testing data, the cue-onset state in the single-rotation RCA
mode predicted response with an area under the receiver operating characteristic curve
(AUC) of 0.638, whereas the pre-cue DRCA state achieved an AUC of 0.612. Combining the
two states increased the AUC to 0.677 (Fig.~\ref{fig:double}D), indicating that the
preparatory DRCA mode contained information complementary to that provided by the RCA
state at cue onset (see Methods for details).

We also detected a double-rotation mode in fMRI recorded during the HCP-YA social task
(Fig.~\ref{fig:double}E). The paradigm alternated between task blocks, in which
participants made a judgment after viewing a video, and fixation blocks. Block-aligned
trajectories occupied different parts of the double rotation: fixation dynamics
circulated around one center, whereas task dynamics circulated around the other
(Fig.~\ref{fig:double}F). Thus, double-rotation modes separated behaviorally defined
epochs in both mouse cortex-wide calcium activity and human fMRI.

\section*{Discussion}

Our results identify broken mirror symmetry as a useful principle for extracting
behaviorally meaningful dynamics from brain-wide recordings. RCA and DRCA consistently
recovered rotational modes across species and recording modalities. These rotations are
defined in a latent phase space, complementing previous observations of brain-wide
rotating waves in physical space~\cite{xu2023spiral,ye2026rotating}. Although these
rotational dynamics do not rely on external signals, their associations with behavior
and learning, together with their alteration in early psychosis, suggest that rotational
flow is a prevalent low-dimensional motif of brain-wide dynamics with functional and
clinical relevance.

An important distinction is the space in which rotation is defined. Ye et al.~\cite{ye2026rotating} detected
topographic waves in which oscillatory phase winds around a center in the physical
cortical sheet, allowing the location, radius, direction and speed of individual wave
events to be measured. By contrast, RCA detects directed
circulation of the brain-wide activity trajectory in a learned latent state space whose
axes are weighted combinations of recorded regions or sensors. It therefore does not
require dense spatial sampling, a predefined frequency band or a spatial wave template,
and can be applied across calcium imaging, fMRI and MEG datasets. A loop in this latent space does
not by itself imply that activity propagates as a physical wave. The two forms of
rotation are nevertheless compatible: a coherent rotating wave could generate a loop in
population state space, whereas latent circulation could also arise from sequential
interactions among distributed regions without forming a spatial vortex. 

Why might rotation recur so broadly? Here we consider two possibilities. First,
cortical circuits may reuse cyclic dynamical motifs that support central pattern
generators and rhythmic movement~\cite{churchland2012neural,marder2001central}. Such motifs could have been adapted to organize more
complex computations, including discrete, non-rhythmic actions. Second, rotation is
among the simplest flows capable of sustaining a nonzero steady-state current and
breaking detailed balance. In a linear system, nonreciprocal coupling between only two macroscopic
degrees of freedom can generate rotational drift, making rotation a natural outcome of
high-dimensional recurrent dynamics. Consistent with this possibility, the trained RNN
developed a spontaneous rotation even though no rotational structure was explicitly
imposed.

The analysis of mouse cortical activity suggests how continuous rotational dynamics can shape a discrete,
non-rhythmic behavioral response. Rotations in motor cortex are well established during individual
reaches, where they have been interpreted as structured population dynamics for
generating movement~\cite{churchland2012neural,russo2018motor,sabatini2024reach}. In our analysis,
the rotation was already present in spontaneous cortex-wide activity before the
cue. Random cue timing therefore sampled different states and directions of this ongoing
flow. Because a dynamical state specifies not only the current activity pattern but also
its direction of evolution, the cue-evoked drive could combine with the intrinsic flow
to make a response-generating trajectory more or less accessible. According to this view, the cue
gates an event-triggered readout of a continuously evolving latent process: neural
activity may cycle, whereas behavior occurs only when the cue and ongoing flow jointly
carry the system toward a response-producing regime. The rotational state thus provides
a coordinate of dynamical readiness, converting structured ongoing dynamics into
trial-to-trial variation in a discrete outcome without requiring rhythmic behavior. The
organization of the cue-evoked drift field further suggests that the intrinsic rotation
may constrain how the response is learned: reinforcement could preferentially strengthen
cue-evoked drives that cooperate with the existing flow, rather than requiring the cortex
to construct a new trajectory that opposes it. This interpretation is consistent with
evidence that animals more readily learn activity patterns within existing neural
manifolds and have difficulty producing trajectories that violate intrinsic dynamics,
suggesting that learning preferentially follows existing neural
structures~\cite{sadtler2014neural,oby2025dynamical}. Our findings extend this
idea from constraints on producing neural trajectories to constraints on learning an
input--response mapping: a new behavior may be acquired by reshaping cue-evoked inputs
within an existing dynamical scaffold rather than by rewriting the intrinsic dynamics.

The reaching benchmarks illustrate where RCA differs from conventional
dimensionality-reduction methods (Fig.~\ref{fig:rca-overview}). In the macaque data,
coherent circulation was evident only in the RCA plane. Both RCA and jPCA revealed
rotation in condition-averaged human activity, but RCA produced a more consistent
individual-trial current. This advantage is expected when trial-specific or other
rotation-independent variation obscures a low-variance rotational mode. Whereas PCA
prioritizes variance and jPCA first restricts the analysis to leading principal
components, RCA directly optimizes the strength and consistency of angular current.
APR is related to the thermodynamic language of irreversibility but should not be
equated with the entropy production rate (EPR). EPR quantifies all detectable
time-reversal asymmetry. Current-based APR isolates the component expressed through a
current that is odd under both time reversal and the chosen spatial transformation.
This construction targets rotational handedness while retaining a quantitative
connection to nonequilibrium probability currents~\cite{crooks1999entropy,
seifert2008stochastic,lynn2021broken}.

Neuronal activity recorded immediately before trial onset (i.e.,~spontaneous) has been shown to nonetheless influence post-trial activity \cite{arieli1996dynamics, Luczak2009Spontaneous}, and even bias behavioral performance \cite{Davis2020Spontaneous}. As discussed above, RCA is a dimensionality reduction technique that can identify within-trial rotational dynamics. Analysis of the mouse cortical activity during the tone-detection/lever-pulling task shows that the position of rotational dynamics before cue onset determined (in part) behavioral response (Fig.~\ref{fig:mouse}F). The simple two-dimensional latent description that RCA provides greatly simplifies the connection from spontaneous to trial-evoked neural activity, compared to past analyses~\cite{arieli1996dynamics,Davis2020Spontaneous}. 

Together, our results support a view in which ongoing brain-wide activity is not merely
background variability around an evoked response. Its directed flow defines coordinates
that shape how an input is converted into action, how rewarded dynamics are reinforced
during learning and how preparatory and task-related states are organized. By optimizing
for broken dynamical symmetry rather than variance, RCA and DRCA provide a route from
high-dimensional recordings to low-dimensional flows with direct behavioral meaning.
Although developed here for brain activity, the framework is broadly applicable to
high-dimensional data and could uncover hidden nonreciprocal dynamics across diverse
physical and biological systems~\cite{fruchart2021nonreciprocal,fruchart2026manybody} such as epigenetic landscapes \cite{Mochulska2025}.

\clearpage
\section*{Online Methods}

\setcounter{figure}{0}
\makeatletter
\renewcommand{\fnum@figure}[1]{{Extended Data Fig.~\thefigure}.~}
\makeatother
\renewcommand{\thefigure}{\arabic{figure}}
\renewcommand{\theHfigure}{ED\arabic{figure}}

\begin{figure}[t]
\centering
\includegraphics[width=0.45\textwidth]{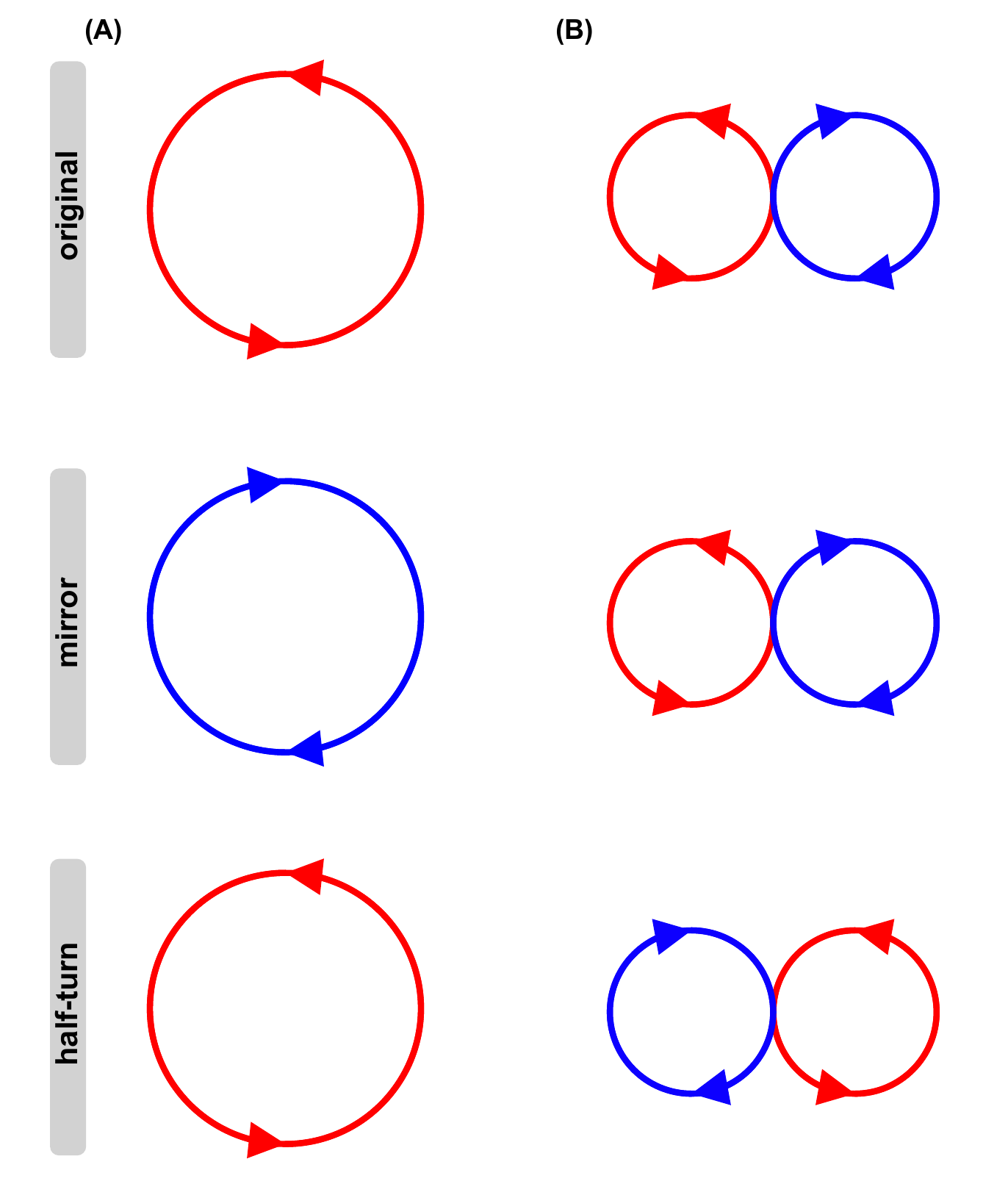}
\caption{\textbf{Symmetry properties of single- and double-rotational currents.}
Red and blue indicate opposite directions of circulation. \textbf{A}, A single
rotational current (top) is converted to the opposite direction by mirror reflection
(middle), thereby breaking mirror symmetry. A half-turn about the center preserves its
circulation pattern (bottom). \textbf{B}, An ideal pair of neighboring counter-rotating
currents (top). Reflection across the midline exchanges the two currents and reverses
their circulation directions, reproducing the original pattern (middle). A half-turn
exchanges their positions without reversing their circulation directions and therefore
produces a different pattern (bottom).}
\label{fig:extended-symmetry}
\end{figure}

\begin{figure*}[t]
\centering
\includegraphics[width=\textwidth]{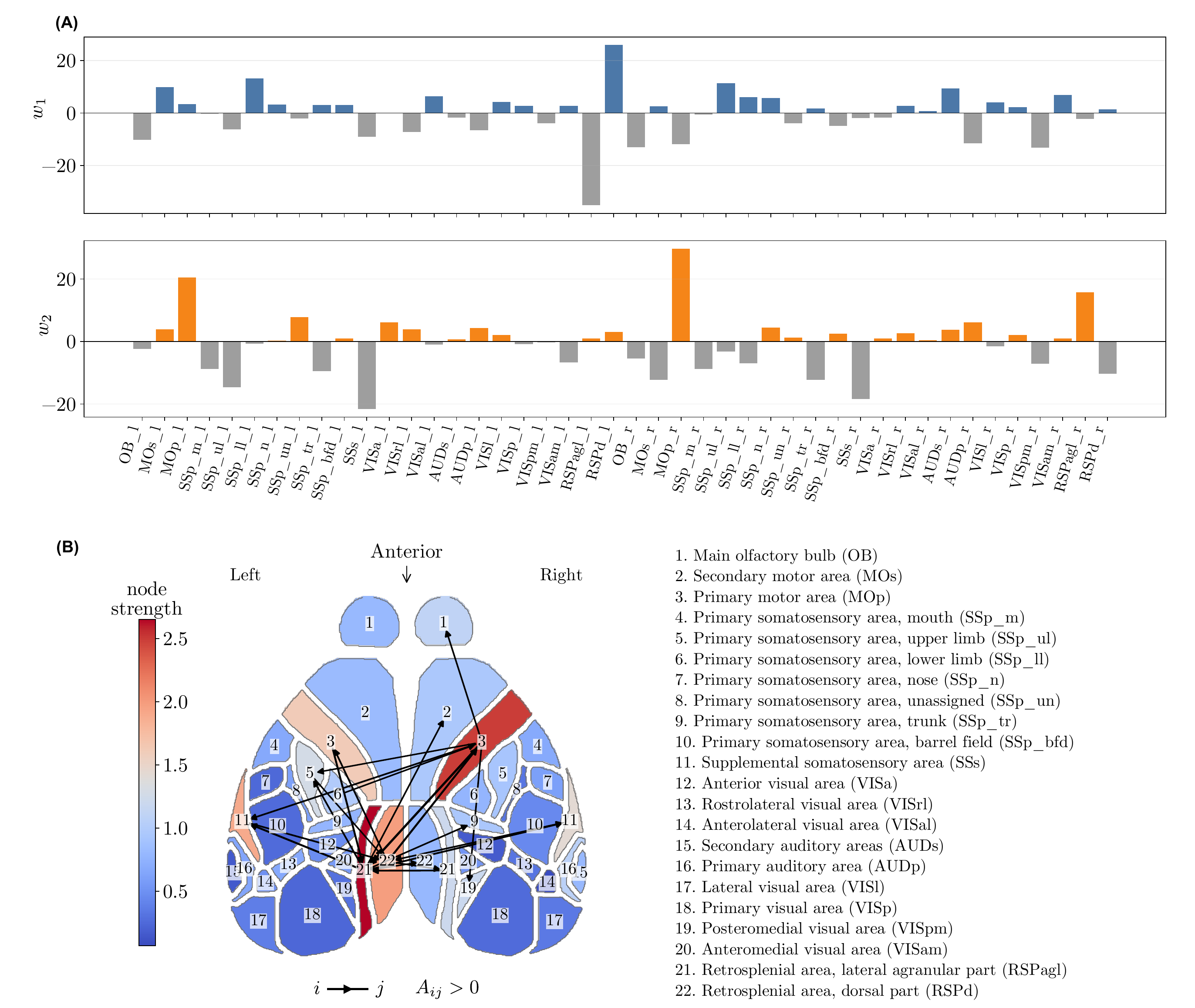}
\caption{\textbf{Anatomical organization of the mouse resting-state rotational
mode.} \textbf{A}, RCA loading vectors $\bm w_1$ and $\bm w_2$ across the 44 recorded
Allen Common Coordinate Framework regions (22 per hemisphere). Blue and orange bars
indicate positive loadings for $\bm w_1$ and $\bm w_2$, respectively, whereas gray bars
indicate negative loadings; suffixes \texttt{l} and \texttt{r} denote the left and right
hemispheres. \textbf{B}, Directed interregional structure implied by the RCA loadings.
We constructed the minimum-Frobenius-norm antisymmetric matrix
$A=2\,\operatorname{sgn}(\omega)(\bm w_1\bm w_2^{\mathsf T}-\bm w_2\bm w_1^{\mathsf T})/
(\lVert\bm w_1\rVert^2+\lVert\bm w_2\rVert^2)$, which represents a unit rotation in
the observed temporal direction in the original region space; $\omega$ is the signed
rotation coefficient in the RCA plane. Region color denotes node strength
$\sum_j|A_{ij}|$, and arrows show the 20 strongest pairwise couplings, oriented such
that $A_{ij}>0$. Numbers identify the regions listed at right; anterior is at the top.
This construction visualizes the interregional coupling pattern implied by the
projection and is not an estimate of anatomical connectivity.}
\label{fig:extended-mode-analysis}
\end{figure*}

\begin{figure}[t]
\centering
\includegraphics[width=0.5\textwidth]{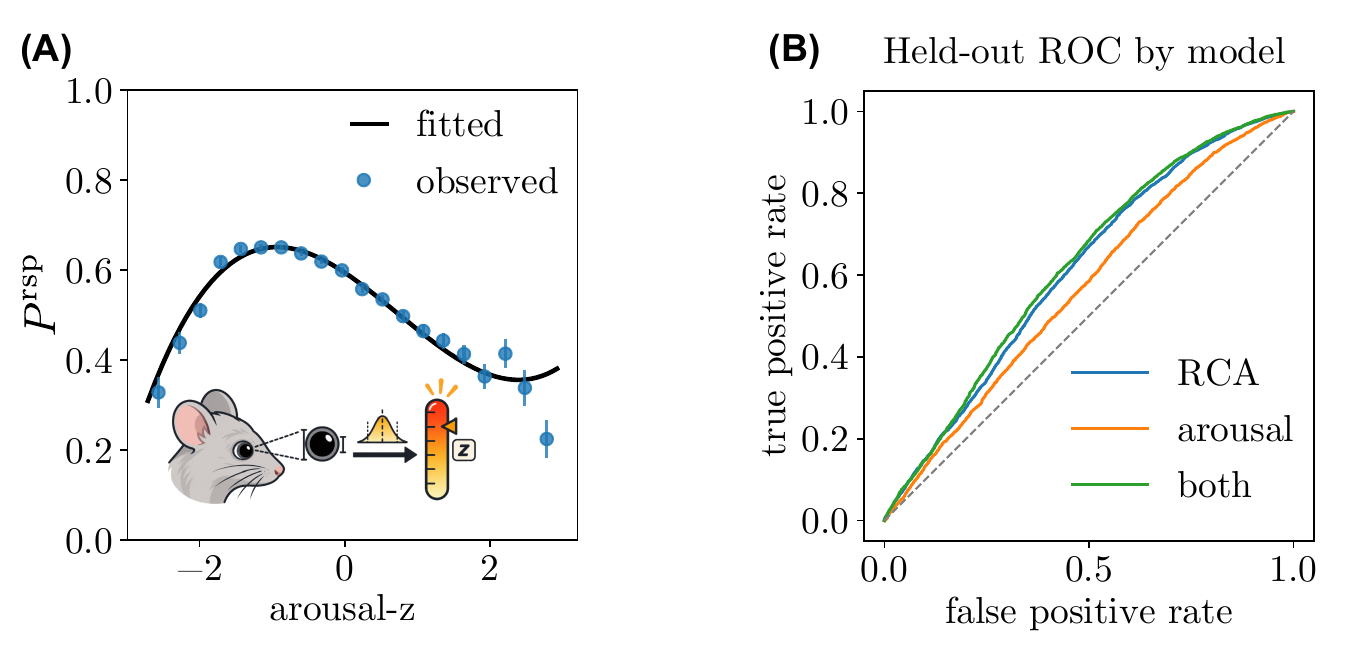}
\caption{\textbf{Response prediction from the RCA state is not explained by
pupil-indexed arousal.} \textbf{A}, Observed response probability $P^{\mathrm{rsp}}$
conditioned on the pupil-derived arousal $z$ score at cue onset. Points show binned
observations from all matched trials, error bars indicate binomial standard errors and
the black curve shows a cubic polynomial fit. This descriptive all-data fit was not
used for held-out evaluation. \textbf{B}, Receiver operating characteristic
curves evaluated on trials from held-out sessions for models using the cue-onset RCA
response score (blue; AUC $=0.611$), a cubic pupil-arousal model
(orange; AUC $=0.535$), or both predictors (green; AUC $=0.611$). Models
were trained on 35,547 trials from 267 sessions and evaluated on 12,272 trials from 89
held-out sessions. The diagonal indicates chance performance.}
\label{fig:extended-pupil}
\end{figure}

\begin{figure*}[t]
\centering
\includegraphics[width=\textwidth]{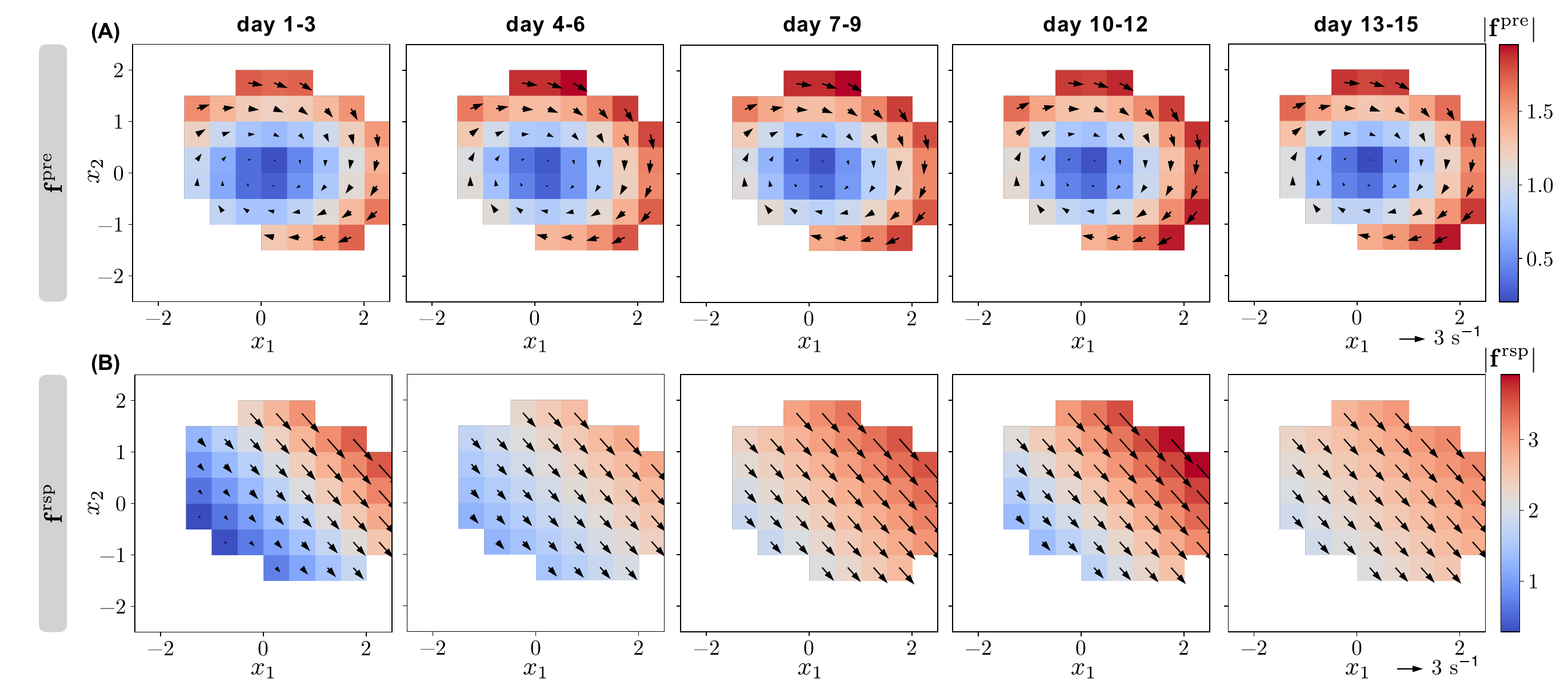}
\caption{\textbf{Intrinsic and cue-evoked drift fields across learning.}
\textbf{A}, Pre-cue intrinsic drift field $\bm f^{\mathrm{pre}}$ estimated independently
in five nonoverlapping three-day windows spanning the 15 task days. Arrows indicate
drift direction and magnitude, and color denotes $|\bm f^{\mathrm{pre}}|$.
\textbf{B}, Cue-evoked response drift field $\bm f^{\mathrm{rsp}}$ estimated in the
same training windows. Arrows indicate drift direction and magnitude, and color denotes
$|\bm f^{\mathrm{rsp}}|$.}
\label{fig:extended-learning}
\end{figure*}

\subsection*{Study design}
This study reanalyzed existing datasets comprising mouse calcium imaging, human fMRI,
human MEG, human intracortical electrophysiology and monkey electrophysiology. No new animals or human participants were
recruited. Sample sizes were determined by the available datasets and were not set by
an a priori power calculation. Ethical approval and consent for the original experiments are described in
the corresponding data publications and repository documentation.

\subsection*{Current-based APR and trajectory-level bounds}
Let $\Gamma=\{\bm x(t):0\leq t\leq\tau\}$ denote a trajectory segment of duration
$\tau$ with path distribution $P$. We seek symmetry breaking carried by temporal
evolution rather than by the stationary distribution alone. We therefore choose a
spatial transformation $\mathcal M$ according to the dynamical pattern of interest and
ask whether trajectory currents distinguish $P$ from its transformed distribution.

To formalize this idea, let $J[\Gamma]$ be a real-valued trajectory current. We require
the current to be additive under concatenation,
$J[\Gamma_1\circ\Gamma_2]=J[\Gamma_1]+J[\Gamma_2]$, and to satisfy
\begin{equation}
J[\mathcal M\Gamma]=-J[\Gamma],
\qquad
J[\mathcal T\Gamma]=-J[\Gamma].
\label{eq:current-parity}
\end{equation}
The first condition makes the current sensitive to breaking of the chosen spatial
symmetry. The second makes it odd under the time-reversal transformation $\mathcal T$.
We denote the class of additive currents satisfying these two parity conditions by
$\mathcal C_{\mathcal M}$. These requirements define an admissible class but do not
select a unique current.

Let $p_J(j)$ denote the distribution of $J[\Gamma]$ for trajectories drawn from $P$.
For each admissible current, we define a current-specific APR and then define the APR
associated with $\mathcal M$ by optimizing over the admissible class:
\begin{equation}
\begin{aligned}
\sigma_{\mathrm{APR}}^{(\mathcal M,J)}
&=\frac{1}{\tau}D_{\mathrm{KL}}
\!\left[p_J(j)\,\Vert\,p_J(-j)\right],\\
\sigma_{\mathrm{APR}}^{\mathcal M}
&=\sup_{J\in\mathcal C_{\mathcal M}}
\sigma_{\mathrm{APR}}^{(\mathcal M,J)}.
\end{aligned}
\label{eq:apr-definition}
\end{equation}
If the dynamics are invariant under $\mathcal M$, the first parity condition requires
$p_J(j)=p_J(-j)$. The same equality follows from time-reversal symmetry and the second
condition. A positive APR therefore requires both spatial-symmetry breaking and temporal
irreversibility. 

A well-chosen current can therefore provide a tight lower bound on the maximal APR. In
practice, both the spatial transformation $\mathcal M$ and the current $J$ should be
chosen according to the dynamical pattern of interest; the construction is nontrivial
only when $\mathcal C_{\mathcal M}$ contains a nonzero current. Extended Data
Fig.~\ref{fig:extended-symmetry} illustrates the two cases studied here. A single
directed rotation breaks reflection symmetry about every axis through its center, making
the angular current in Eq.~\ref{eq:angular-current} a natural choice because it changes
sign under each such reflection. By contrast, an ideal pair of counter-rotating vortices
can preserve the reflection that exchanges the two vortices while breaking half-turn
symmetry about $\bm c$. The modified current $J^*=(x_1-c_1)J$ detects this pattern: it
changes sign under the half-turn $\bm x\mapsto2\bm c-\bm x$ but remains invariant under
the reflection $x_1-c_1\mapsto-(x_1-c_1)$.

The current-based APR is bounded by the corresponding asymmetries of the full
trajectory distribution. Define the transformed distributions
$P^{\mathcal M}=\mathcal M_{\#}P$ and $P^{\mathcal T}=\mathcal T_{\#}P$. The
data-processing inequality~\cite{cover2006elements} and Eq.~\ref{eq:current-parity}
give
\begin{equation}
\begin{aligned}
0\leq\sigma_{\mathrm{APR}}^{(\mathcal M,J)}
&\leq\sigma_{\mathrm{APR}}^{\mathcal M}\\[-2pt]
&\leq\min\!\left\{
\frac{D_{\mathrm{KL}}(P\Vert P^{\mathcal M})}{\tau},
\frac{D_{\mathrm{KL}}(P\Vert P^{\mathcal T})}{\tau}
\right\}.
\end{aligned}
\label{eq:apr-path-bounds}
\end{equation}
The temporal path-space divergence is a standard expression for EPR
\cite{crooks1999entropy,kawai2007dissipation,seifert2008stochastic}. It has also been
used to quantify temporal irreversibility in large-scale brain dynamics
\cite{lynn2021broken,cruzat2023temporal}. With complete physical variables and the
appropriate stochastic-thermodynamic assumptions, the temporal path-space divergence is
the thermodynamic EPR. Neural recordings, however, provide coarse-grained observables,
and current-specific APR retains only the temporal asymmetry expressed through the
selected current. We therefore interpret APR as a statistical measure of dynamical
symmetry breaking and do not assign it a thermodynamic meaning a priori.

For an admissible accumulated current $J_\tau$ whose distribution is Gaussian, with
mean $\overline{J_\tau}$ and variance $\Delta J_\tau$, the current-specific APR is
exactly
\begin{equation}
\sigma_{\mathrm{APR}}^{(\mathcal M,J)}
=\frac{1}{\tau}\frac{2\overline{J_\tau}^{\,2}}{\Delta J_\tau}
\leq\sigma_{\mathrm{APR}}^{\mathcal M}.
\label{eq:gaussian-current-apr}
\end{equation}
The moment ratio in Eq.~\ref{eq:gaussian-current-apr} is the objective
$\widehat\sigma$ used in
Eq.~\ref{eq:apr-estimator}. Thus, under the Gaussian assumption, RCA maximizes a
current-specific lower bound on the transformation-specific APR. Its
mean-current-to-variance form parallels thermodynamic uncertainty relations
\cite{barato2015thermodynamic,gingrich2016dissipation}. For a general, non-Gaussian
current distribution, the involution uncertainty relation gives
\begin{align}
\sigma_{\mathrm{APR}}^{\mathcal M}
&\geq\sigma_{\mathrm{APR}}^{(\mathcal M,J)}
\geq\frac{1}{\tau}B(u),
&u&=\frac{\Delta J_\tau}{\overline{J_\tau}^{\,2}}, \nonumber\\
B(u)&=\frac{2}{\sqrt{1+u}}
\operatorname{artanh}\!\left(\frac{1}{\sqrt{1+u}}\right).&&
\label{eq:universal-apr-bound}
\end{align}
For weak asymmetry, $u\gg1$ and
$B(u)=2/u-4/(3u^2)+O(u^{-3})$. Thus, $\widehat\sigma$ is the leading
quadratic approximation to this nonlinear lower bound.

For empirical estimation, we used one-step trajectory segments after all temporal
binning or downsampling. Thus, for adjacent samples $\bm x_k$ and $\bm x_{k+1}$
separated by $\tau=\Delta t$, the accumulated angular current was discretized as
\begin{equation}
\begin{aligned}
J_{\Delta t,k}={}&(x_{1,k}-c_1)(x_{2,k+1}-x_{2,k})\\
&-(x_{2,k}-c_2)(x_{1,k+1}-x_{1,k}).
\end{aligned}
\label{eq:discrete-angular-current}
\end{equation}
The mean and variance in Eq.~\ref{eq:apr-estimator} were computed over all valid
within-trajectory transitions used to fit a given projection; transitions were never
formed across trial, epoch, run or session boundaries. The effective values of
$\Delta t$ were $1/30$~s for mouse resting-state calcium imaging and human MEG,
0.72~s for human fMRI, 50~ms for human reaching and 10~ms for macaque reaching. Mouse
task data used three-frame temporal coarse-graining for DRCA, giving
$\Delta t=0.10$~s. The prefactor $1/\tau$ expresses the segment-level statistic as a
rate. Estimates obtained at different temporal resolutions need not be numerically
identical because temporal coarse-graining changes the current distribution.

\begin{figure*}[t]
\centering
\includegraphics[width=\textwidth]{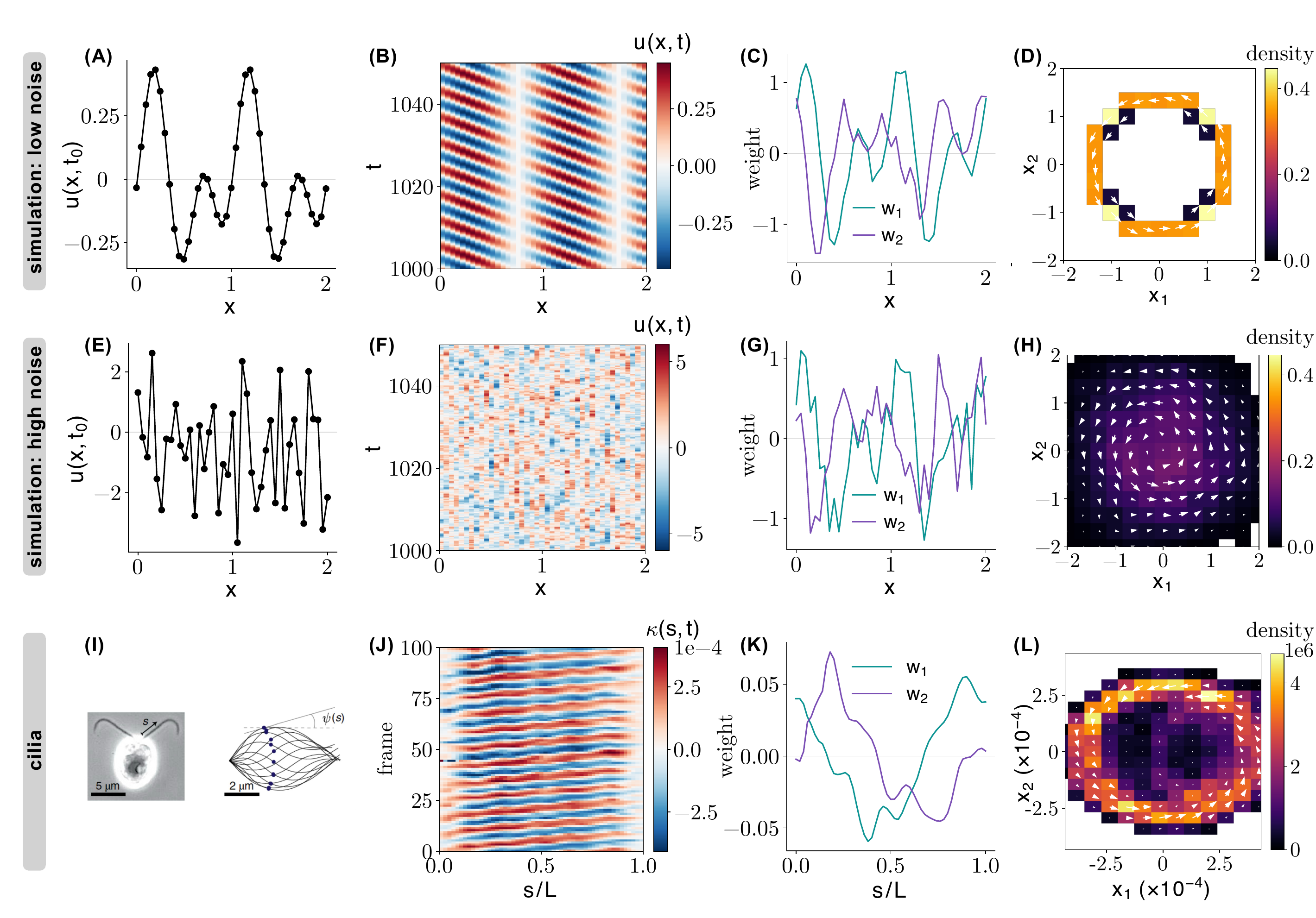}
\caption{\textbf{RCA beyond neuroscience: recovering rotational modes from noisy traveling patterns and
ciliary beating.} \textbf{A--D}, A simulated two-mode traveling pattern with weak
Ornstein--Uhlenbeck noise ($D=10^{-7}$). \textbf{A}, Spatial profile $u(x,t_0)$ at one
time point. \textbf{B}, Space--time evolution of $u(x,t)$; diagonal bands indicate
traveling waves. \textbf{C}, Spatial weights $w_1$ and $w_2$ of the two RCA
coordinates. \textbf{D}, Probability density (color) and probability flux (white
arrows) in the standardized RCA plane. \textbf{E--H}, Corresponding analysis after
increasing the noise strength to $D=0.1$. Although the traveling pattern is obscured in
the instantaneous profile (\textbf{E}) and space--time map (\textbf{F}), RCA recovers
structured spatial weights (\textbf{G}) and a coherent circulating probability flux
(\textbf{H}). \textbf{I--L}, RCA analysis of experimentally measured ciliary beating
from Geyer et al.~\cite{geyer2022ciliary}.
\textbf{I}, Example cilium and centerline representation used to measure tangent angle
$\psi(s)$ and curvature along arclength $s$. \textbf{J}, Curvature $\kappa(s,t)$ along
normalized arclength $s/L$ over 100 consecutive frames from an example recording.
\textbf{K}, Mean-subtracted RCA weights $w_1$ and $w_2$ along the cilium.
\textbf{L}, Probability density (color) and probability flux (white arrows) in the RCA
plane, estimated from concatenated 500-frame curvature segments.}
\label{fig:extended-cycle}
\end{figure*}

To test whether the current-based RCA objective could robustly detect hidden rotational degrees of
freedom, we applied RCA to simulated traveling patterns and experimental ciliary
beating data from Geyer et al.~\cite{geyer2022ciliary} (Extended Data
Fig.~\ref{fig:extended-cycle}). We generated the simulated
pattern from two coupled complex amplitudes, $A_1(t)$ and $A_2(t)$, obeying
\begin{align}
\dot A_1 & = \mu_1 A_1-A_2-\alpha |A_1|^2A_1-\beta |A_2|^2A_1, \nonumber\\
\dot A_2 & = \mu_2 A_2+\epsilon A_1-\gamma |A_1|^2A_2-\delta |A_2|^2A_2,
\label{eq:traveling-pattern-amplitudes}
\end{align}
with $\mu_1=\mu_2=0.1$ and
$\alpha=\beta=\gamma=\delta=\epsilon=1$. The observed spatial field
was
\begin{equation}
u(x,t)=\operatorname{Re}\!\left[A_1(t)e^{\mathrm{i}kx}
+A_2(t)e^{2\mathrm{i}kx}\right]+\zeta(x,t),
\label{eq:traveling-pattern-field}
\end{equation}
where $k=2\pi$ and $\zeta(x,t)$ was an Ornstein--Uhlenbeck noise field,
\begin{align}
\partial_t\zeta(x,t)&=-\zeta(x,t)+\xi(x,t), \nonumber\\
\left\langle\xi(x,t)\xi(x',t')\right\rangle
&=2D\,\delta(x-x')\delta(t-t').
\label{eq:traveling-pattern-noise}
\end{align}
An interactive browser-based simulation, with adjustable model parameters and noise
strength, is available online~\cite{chen2026hiddenrotation}. 
In steady state, the model produces a limit-cycle traveling pattern that can be described
by two spatial components whose amplitudes circulate in a reduced
plane~\cite{fruchart2021nonreciprocal}. For the simulated pattern, RCA recovered
these spatial components and the associated rotational probability current when the
wave was readily visible (Extended Data Fig.~\ref{fig:extended-cycle}A--D). After the
noise strength was increased such that the traveling pattern was obscured in both the
instantaneous spatial profile and the space--time map, RCA still identified structured
projection weights and coherent circulation (Extended Data
Fig.~\ref{fig:extended-cycle}E--H). Applied to measurements of ciliary curvature, RCA
likewise recovered two spatial modes and a circulating current associated with the
propagating bend along the cilium (Extended Data
Fig.~\ref{fig:extended-cycle}I--L). These examples show that the APR objective can
recover latent rotational coordinates generated by propagation even when rotation is
not directly visible in the high-dimensional observations.

\begin{figure}[t]
\centering
\includegraphics[width=0.45\textwidth]{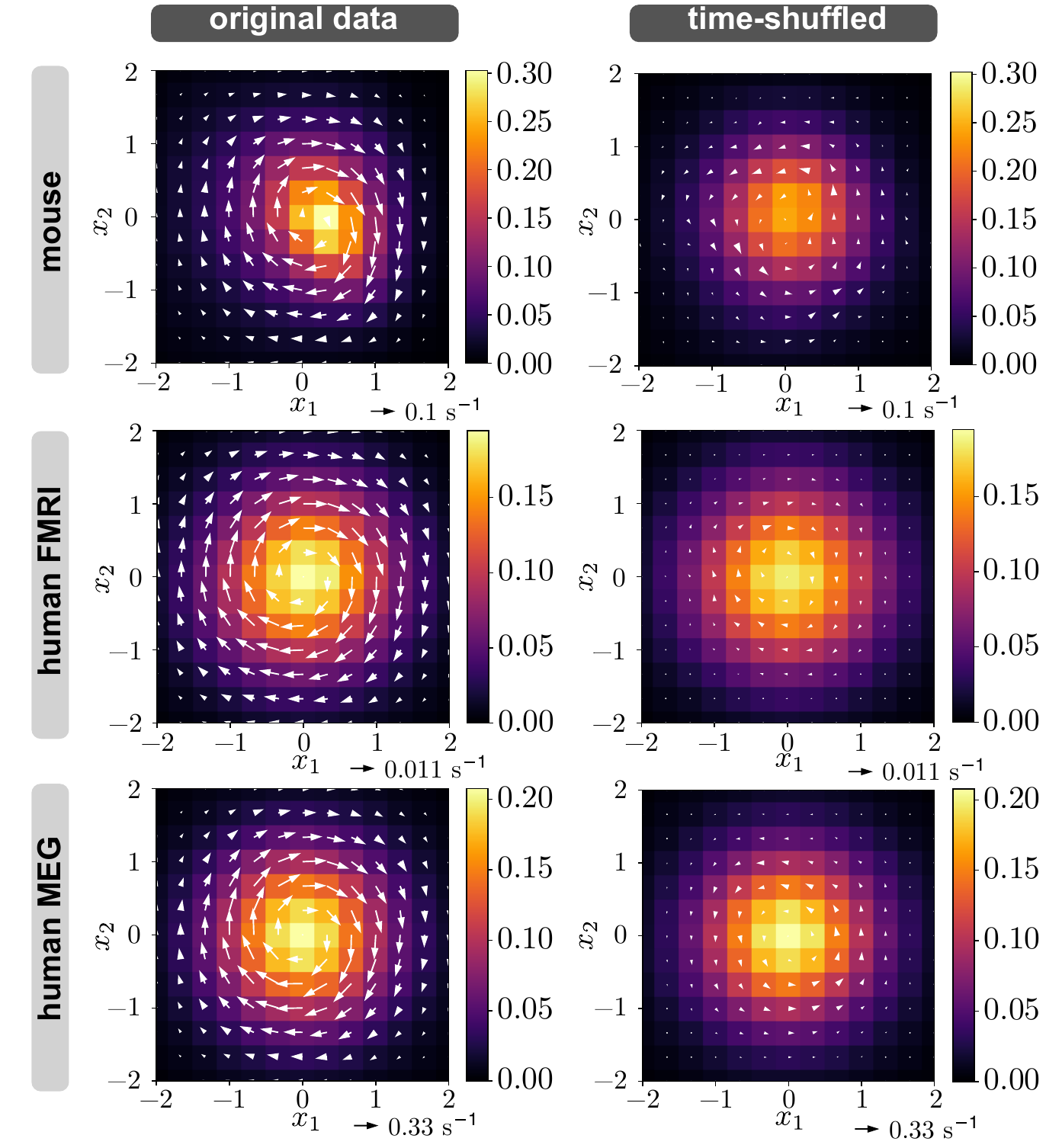}
\caption{\textbf{Temporal shuffling strongly suppresses rotational probability currents.}
Probability density (color) and probability flux (white arrows) are shown in RCA planes
for mouse resting-state calcium imaging (top), HCP-YA resting-state fMRI (middle) and
HCP-YA resting-state MEG (bottom). The left column shows the original data. For the
control in the right column, time points were independently permuted within each
trajectory and RCA was relearned from the shuffled data. This procedure preserves the
set of instantaneous multivariate states within every trajectory while destroying
their temporal order. Coherent circulation was present in each original dataset but
was strongly attenuated after shuffling, indicating that the identified rotational
currents reflect ordered temporal transitions rather than the stationary distribution
of states or optimization over the projection plane.
Scale arrows indicate the probability-flux scale.}
\label{fig:extended-shuffle}
\end{figure}

Because angular current is also odd under time reversal, the same data-processing argument
relates its directional asymmetry to EPR; its population mean vanishes for a
time-reversal-symmetric process. With finite data, however,
sampling fluctuations and optimization over projection planes can produce an apparent
nonzero rotation. To test whether the observed rotations depended on temporal ordering,
we independently permuted the time points within every trajectory in the mouse calcium,
human fMRI and human MEG datasets. This
procedure preserved the observed multivariate states while destroying the transitions
between successive states; we then relearned the RCA plane from each shuffled dataset.
The original data showed coherent circulation in all three modalities, whereas the
shuffled controls contained only much weaker, disorganized currents (Extended Data
Fig.~\ref{fig:extended-shuffle}). Thus, the rotations in the original data depend on
ordered temporal transitions rather than only on the distribution of instantaneous
states.

\subsection*{Normalization and orientation of two-dimensional modes}
Across all datasets, we treated a two-dimensional linear mode first as a subspace,
independent of the scaling of its two weight vectors. Let
$V=[\bm v_1,\bm v_2]\in\mathbb{R}^{N\times2}$ contain the vectors returned by a
dimensionality-reduction method. We computed the reduced QR decomposition $V=QR$,
where $Q^{\mathsf T}Q=I_2$. The columns of $Q$ form an orthonormal basis for the same
plane as the original vectors, thereby removing arbitrary differences in their lengths
and mutual angle. This orthonormalization was performed in the feature space to which
the mode was applied, after the dataset-specific preprocessing described below.

We used the orthonormal basis to calculate the fraction of variance explained by the
plane,
\begin{equation}
\frac{\operatorname{tr}(Q^{\mathsf T}CQ)}{\operatorname{tr}(C)},
\end{equation}
where $C$ is the covariance matrix of the preprocessed population activity. This
quantity depends only on the plane and not on the choice of axes within it. The
within-plane basis is otherwise non-unique and was oriented according to the mode being
analyzed. For an RCA mode, we diagonalized the projected covariance
$Q^{\mathsf T}CQ$ and chose the first axis along the eigenvector with the larger
eigenvalue. For a DRCA mode, whose objective is not invariant under arbitrary
within-plane rotations, the first axis was chosen to maximize
Eq.~\ref{eq:double}. 

Finally, we standardized the scale of the two oriented coordinates. If
$\bm q_i$ denotes an oriented unit vector and
$s_i^2=\bm q_i^{\mathsf T}C\bm q_i$ is the variance of its projection, the weight used
to report the mode was
\begin{equation}
\bm w_i=\frac{\bm q_i}{s_i},
\qquad x_i=\bm w_i^{\mathsf T}\bm y,
\end{equation}
where $\bm y$ is the preprocessed population state, so that
$\operatorname{Var}(x_i)=1$ for both coordinates. Thus, $Q$ defines the
orthonormal plane and its explained variance, whereas $\bm w_1$ and $\bm w_2$ define
the unit-variance coordinates used in subsequent analyses and figures. This convention
placed modes from different methods and datasets on a common scale, including the
response-gradient comparison in Fig.~\ref{fig:mouse}H.

\subsection*{Mouse wide-field calcium imaging}
We used the processed BraiDyn-BC dataset~\cite{kondo2025multimodal}. The dataset contains
recordings from 25 head-fixed VGluT1-Cre $\times$ Ai162 mice expressing GCaMP6s in
excitatory cortical neurons, obtained during rest and across 15 daily sessions of a
cued lever-pull task.
A 10-kHz tone initiated a 1-s response window, and a successful pull triggered delivery
of a 4-$\mu$l water reward. Cue timing included a random pre-cue delay. We analyzed
the processed signals averaged within 44 cortical regions registered to the Allen Common
Coordinate Framework (22 regions per hemisphere). We used these released regional
traces directly and did not reprocess the raw wide-field movies. ROI ordering was
required to be identical across files, and timestamps were retained to align neural
activity to behavioral events. The effective sampling rate was approximately 30~Hz.

The resting-state analysis included 98 sessions from all 25 mice. Each session
contained 18,000 samples from 44 regions, corresponding to approximately 600~s of
recording. All 98 traces had the complete expected duration and dimensionality, so no
resting-state session was excluded. No temporal coarse-graining was applied in the
analyses reported here.

The task dataset contained 364 sessions from all 25 mice, spanning task days 1--15
(13--15 sessions per mouse). Each session contained 54,000 samples, corresponding to
approximately 1,800~s. The behavioral table contained 48,549 trials with finite cue
times. For cue-aligned analyses, we extracted 240-frame windows comprising 120 pre-cue
and 120 post-cue frames (approximately 4~s on either side of cue onset). A trial was
retained only if the complete $240\times44$ window lay within the valid trace and all
values were finite. This procedure retained 48,342 trials from all 364 sessions and all
25 mice: 28,760 trials with a recorded pull onset and 19,582 without a recorded pull.
We excluded 207 trials because the requested window crossed a recording boundary (107
before the start and 100 after the end); no trials were excluded because of nonfinite
values, an unexpected window shape or a missing processed session.

\subsection*{Anatomical representation of the mouse RCA mode}
To visualize the anatomical organization of the resting-state rotational mode, we
examined the two RCA loading vectors, $\bm w_1,\bm w_2\in\mathbb{R}^{44}$, that define
the latent coordinates $x_1=\bm w_1^{\mathsf T}\bm y$ and
$x_2=\bm w_2^{\mathsf T}\bm y$ from the regional activity vector $\bm y$. The signed
loading of each Allen Common Coordinate Framework region on the two axes is shown in
Extended Data Fig.~\ref{fig:extended-mode-analysis}A. Because the orientation of the
coordinate axes within an identified plane is arbitrary, the two loading vectors should
be interpreted jointly as specifying a rotational subspace rather than as two
independently meaningful anatomical components.

We summarized the directed regional organization implied by this plane using a
rank-two antisymmetric matrix. Let $s$ denote the rotation direction such that $s=1$ for clockwise rotation and $s=-1$ for counterclockwise rotation. 
We defined the corresponding unit-magnitude generator in regional space as
\begin{equation}
A=\frac{2\,s\left(\bm w_1\bm w_2^{\mathsf T}
-\bm w_2\bm w_1^{\mathsf T}\right)}
{\lVert\bm w_1\rVert^2+\lVert\bm w_2\rVert^2}.
\label{eq:mode-coupling-matrix}
\end{equation}
This construction embeds a unit rotational generator in the subspace spanned by the
two RCA loading vectors while introducing no coupling in the orthogonal complement.
It is antisymmetric, $A_{ij}=-A_{ji}$, and therefore assigns an orientation to each
pairwise contribution without introducing a reciprocal component. For each region, we
defined node strength as $s_i=\sum_j|A_{ij}|$. In Extended Data
Fig.~\ref{fig:extended-mode-analysis}B, region color represents $s_i$, and the 20
largest off-diagonal elements by absolute magnitude are displayed as arrows, oriented
from region $i$ to region $j$ when $A_{ij}>0$. Region locations were taken from the
Allen cortical map and displayed separately for the two hemispheres. Because $A$ is
constructed solely from the RCA projection weights, it visualizes the directed
interregional organization implied by the latent rotational mode; it should not be
interpreted as a measurement of anatomical connectivity, synaptic coupling or causal
influence between regions.

Inspection of the ranked matrix elements showed that the strongest antisymmetric
coupling linked right primary motor cortex (3) to left lateral agranular retrosplenial
cortex (21). Four of the seven strongest couplings formed the
closed sequence right primary motor cortex (3) $\rightarrow$ left lateral agranular
retrosplenial cortex (21) $\rightarrow$ left supplemental somatosensory cortex (11)
$\rightarrow$ left dorsal retrosplenial cortex (22) $\rightarrow$ right primary motor
cortex (3). The displayed structure is therefore dominated by a long-range,
interhemispheric loop spanning sensorimotor and retrosplenial regions, rather than by
interactions restricted to neighboring cortical areas. This organization
provides a possible dynamical scaffold linking ongoing cortical state to subsequent
sensorimotor processing. 

\subsection*{Human fMRI}
We analyzed cortical fMRI data from the HCP Young Adult S1200 release using the cohort
and preprocessed time series described previously~\cite{barch2013function,lynn2021broken}.
Cortical data were registered to fsLR-32k space using MSM-All and parcellated into 100
regions. Five component-based noise-correction (CompCor) components derived from
ventricular and white-matter masks and 12 detrended HCP motion regressors were removed.
The mean global signal was regressed out, after which the time series were band-pass
filtered between 0.009 and 0.08 Hz. Frames with framewise displacement greater than
0.2 mm or DVARS greater than 75 were treated as outliers, and sessions in which more
than 50\% of frames were outliers were excluded. Participants were retained only if all
selected scans survived quality control, yielding 590 healthy adults.

For the resting-state analysis, we used the REST1 LR and RL runs (repetition time,
0.72 s) and divided each run into nonoverlapping 50-frame trajectories. For the social
task analysis, we used the complete LR and RL runs (274 frames per run), yielding 1,180
participant-run trajectories from 590 participants for mode estimation. HCP task labels
were not used to identify the projection and were introduced only afterward to align and
average task and fixation trajectories. For this block-aligned analysis, participants
whose block order differed from the majority order were excluded separately for each
phase-encoding direction (3 from LR and 2 from RL), leaving 587 LR runs and 588 RL runs
from 589 unique participants.

\subsection*{Human MEG}
We analyzed resting-state MEG recordings from the HCP Young Adult S1200
release~\cite{larsonprior2013adding}. Data were acquired with a whole-head MAGNES 3600
system containing 248 magnetometer channels, at a sampling rate of 2,034.51~Hz, in up
to three approximately 6-min runs per participant. We used the released channel-level
\texttt{rmegpreproc} outputs. In the HCP pipeline, bad channels and time segments were
removed, independent components classified as artifacts, including ocular and cardiac
components, were projected out, and the cleaned data were downsampled to approximately
508.63~Hz and segmented into artifact-clean 2-s epochs.

We used 267 preprocessed runs and constructed a common sensor space by selecting
the 238 channels present in at least 95\% of runs (at least 254 of 267). We then retained
only runs containing all 238 selected channels, without channel imputation. This
procedure yielded 140 runs from 71 participants, comprising 19,960 artifact-clean
epochs. Within each epoch, we averaged neighboring samples in consecutive
$1/30$-s bins, producing an independent $60\times238$ trajectory sampled at 30~Hz; no
transitions were formed across epoch boundaries. Only trajectories with finite values
and the common channel ordering were analyzed.

\subsection*{HCP Early Psychosis fMRI}
We analyzed the derived AAL116-parcellated resting-state fMRI time series distributed
with the HCP Early Psychosis (HCP-EP) study~\cite{hancock2023metastability}. The dataset
comprised 53 healthy controls and 82 participants with non-affective psychosis. Each
participant completed four resting-state runs over two consecutive days, with 410 frames
per run and a repetition time of 0.72 s. The source images were processed using the HCP
minimal preprocessing pipeline and denoised using ICA-FIX; regional time series were
band-pass filtered between 0.01 and 0.08 Hz. We retained all four runs, yielding 212
control runs and 328 non-affective-psychosis runs.

\begin{figure}[t]
\centering
\includegraphics[width=0.5\textwidth]{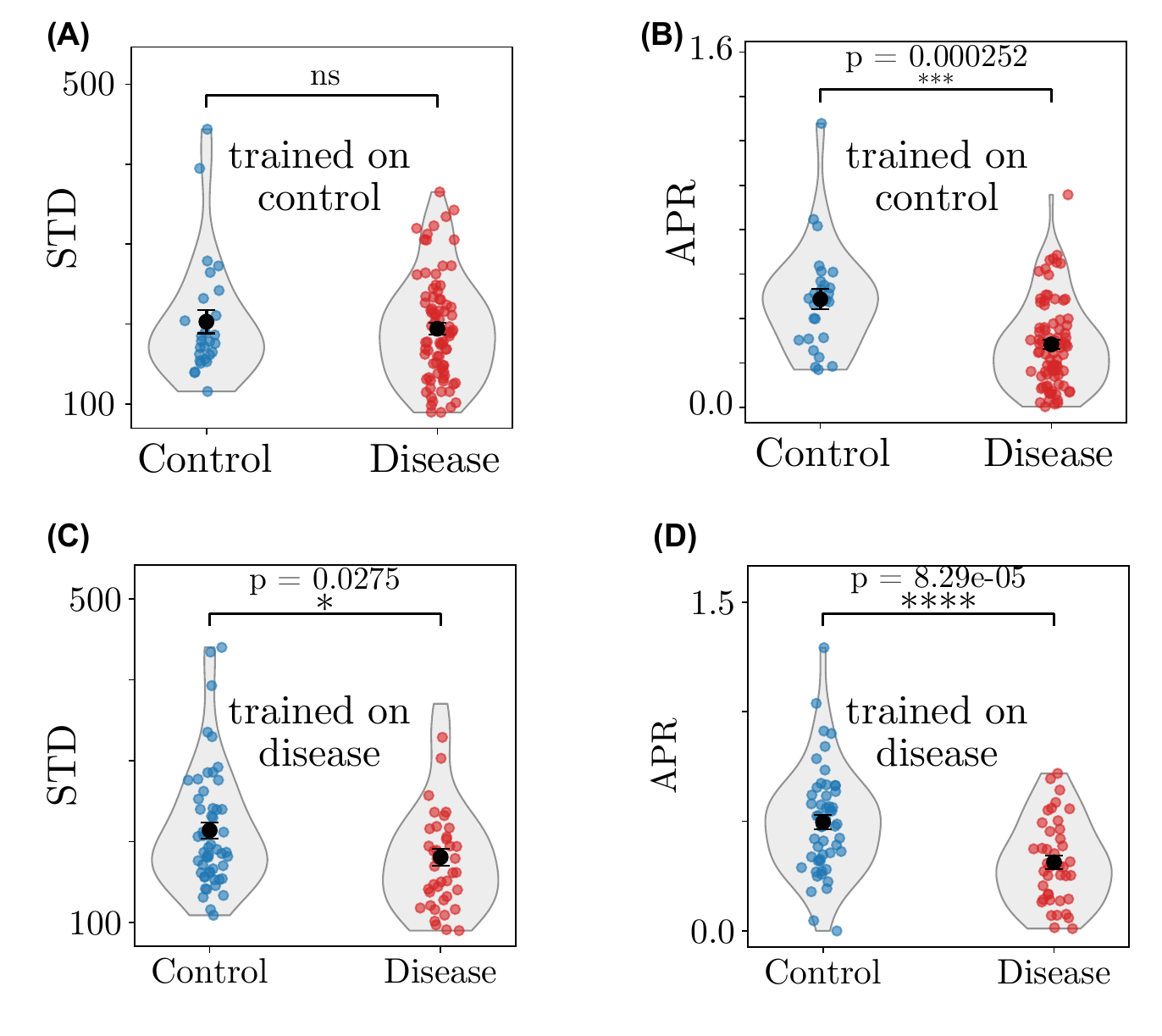}
\caption{\textbf{Split-sample analyses of reduced rotational irreversibility in early
psychosis.} \textbf{A,B}, Control-trained analysis. RCA and PCA projections were
learned from 26 randomly selected control participants and evaluated in the remaining
27 controls and all 82 participants with non-affective psychosis. Participant-level
standard deviation (STD) along the learned leading principal component did not differ
between groups (\textbf{A}; $P=0.592$), whereas APR in the learned RCA plane was lower
in participants with psychosis (\textbf{B}; $P=0.000252$). \textbf{C,D},
Disease-trained analysis. Projections were learned from 41 randomly selected
participants with psychosis and evaluated in the remaining 41 participants and all 53
controls. PCA standard deviation showed only a modest, single-asterisk difference
(\textbf{C}; $P=0.0275$), whereas APR showed a more pronounced group difference
(\textbf{D}; $P=8.29\times10^{-5}$). Points indicate participants, violins show the group distributions,
and black symbols and error bars indicate means $\pm$ s.e.m. Group comparisons used
two-sided Welch's $t$-tests; ns, not significant.}
\label{fig:extended-hcpep}
\end{figure}

In the primary analysis, RCA and PCA projections were learned exclusively from
healthy-control data. The projections were fitted using all 53 controls, and
participant-level APR and PCA variability were calculated by combining the four runs
from each participant; PCA variability
was defined as the standard deviation along the leading principal component. In the
control-trained split-sample analysis, 26 randomly selected controls were used for projection
learning, while the remaining 27 controls and all 82 participants with non-affective
psychosis were held out for evaluation. Group differences were assessed using two-sided
Welch's $t$-tests, with participants as the independent analysis units. In this
control-trained analysis, PCA variability did not differ between groups ($P=0.592$),
whereas APR remained lower in participants with non-affective psychosis
($P=2.52\times10^{-4}$; Extended Data Fig.~\ref{fig:extended-hcpep}A,B).

We performed a complementary disease-trained analysis to test whether this result
depended on learning the projections from control data. We randomly divided the 82
participants with non-affective psychosis into two equal groups, learned the RCA and
PCA projections from 41 participants, and evaluated them in the remaining 41
participants and all 53 controls. In this analysis, PCA variability was only modestly
lower in the held-out psychosis group ($P=0.0275$; one asterisk), whereas APR showed a
more significant reduction ($P=8.29\times10^{-5}$; Extended Data
Fig.~\ref{fig:extended-hcpep}C,D). Thus, reduced APR was detected irrespective of which
group supplied the data used to learn the rotational plane.

\subsection*{Reaching-data benchmark}
We benchmarked RCA against jPCA using the macaque and human motor-cortical recordings
shown in Fig.~\ref{fig:rca-overview}C--N. We performed two complementary analyses. First,
following the conventional jPCA analysis, we fitted each method to activity averaged
across repeated trials within each reach condition~\cite{churchland2012neural}. Second,
we fitted each method directly to the same individual-trial activity pooled across reach
conditions, without averaging trials before dimensionality reduction. The first analysis
tested whether RCA recovered the established condition-averaged rotational structure;
the second tested whether each method could isolate circulation in the presence of
trial-specific neural variability.

For the human analysis, we used motor-cortical activity
recorded from participant C1 during an active center-out reaching task~\cite{tian2026firing}. C1 was male,
55--60 years old at implantation, and had a chronic C4-level, ASIA D spinal cord injury.
The original study was conducted under a US Food and Drug Administration
Investigational Device Exemption, was approved by the Institutional Review Boards of
the Universities of Pittsburgh and Chicago, and was registered at ClinicalTrials.gov
(NCT01894802). The participant provided informed consent before study procedures.

Neural activity was recorded from two $4\times4$-mm microelectrode arrays implanted in
motor cortex, each with 96 active electrode shanks. Voltages were sampled at 30~kHz,
high-pass filtered at 750~Hz, thresholded at $-4.5$ times the root-mean-square voltage
and sorted offline. During the task, the participant's arm rested on a low-friction
forearm support, and eight targets were arranged around a circle 30~cm in diameter.
Each trial comprised five 1-s phases indicated by auditory tones: target cue,
movement, hold, return and a final wait. The target direction was presented as text at
cue onset, and the participant moved the arm toward the target after the movement
tone. Participant C1 completed 11 blocks, with ten repetitions of each target per
block, for a total of 880 reaches. For our analysis, spike trains were aligned to the
first auditory tone (cue onset), binned at 50~ms across the full 0--5-s trial and
converted to firing rates. The firing-rate traces were smoothed with a Gaussian kernel
with a standard deviation of 100~ms. For the RCA--jPCA comparison, we retained activity
from the second tone (movement instruction; median latency from cue onset, 1.01~s) to the
end of the trial. This interval contained 80 bins, with centers spanning 1.025--4.975~s
relative to cue onset.

For the macaque analysis, we used the publicly available motor-cortical activity of
monkey Jenkins from the delayed-reaching dataset of Churchland and
colleagues~\cite{churchland2012neural}. Spike counts from 137 units were aligned to
movement onset, binned at 10~ms from $-250$ to $450$~ms and smoothed with a Gaussian
kernel with a standard deviation of 20~ms. The 2,295 analysis trials were grouped into
36 maze conditions, each containing at least 20 trials. Projection planes were fitted
using the full analysis interval.

For both datasets, we first calculated the mean firing-rate trajectory for each reach
condition. For each unit, we calculated the range of its condition-averaged activity
across conditions and time. Both the condition averages and every individual-trial
firing-rate trace were divided by this range plus a 5-Hz soft-normalization constant.
At each time point, we then subtracted the mean normalized activity across conditions.
This common transformation ensured that the condition-average and individual-trial fits
used the same unit scaling and time-dependent baseline.

For the condition-average analysis, RCA and jPCA were fitted to the resulting
condition-by-time-by-unit tensor. RCA learned two projection vectors by maximizing the
angular-current objective in Eq.~\ref{eq:apr-estimator}. For jPCA, the same tensor was
first reduced to six principal components, a skew-symmetric linear dynamical system was
fitted to the temporal derivative and the complex eigenplane with the highest rotational
frequency was retained~\cite{churchland2012neural}. The fitted condition-average
trajectories are shown for macaque (Fig.~\ref{fig:rca-overview}C,D) and human
(Fig.~\ref{fig:rca-overview}I,J) reaching.

For the individual-trial analysis, RCA and jPCA were refitted from scratch using all
preprocessed individual trials pooled across reach conditions. The methods received
identical trial-level inputs but selected planes using their respective angular-current
and skew-symmetric-dynamics objectives. For visualization, we projected each trial into
the corresponding fitted plane and then averaged the projected trajectories within each
condition. These post-fit condition averages are shown for macaque
(Fig.~\ref{fig:rca-overview}E,F) and human (Fig.~\ref{fig:rca-overview}K,L) reaching;
colors distinguish conditions, and dots mark trajectory onset.

We quantified trial-level circulation in the individual-trial-fit planes for macaque
maze condition 1 (Fig.~\ref{fig:rca-overview}G,H) and the representative human
bottom-target condition (Fig.~\ref{fig:rca-overview}M,N). For each method and dataset,
the fitted plane was expressed as an orthonormal basis before projection. We estimated
the empirical center and mean squared radius from all projected samples in the selected
condition. The normalized cumulative angular current was then calculated from finite
differences using Eq.~\ref{eq:normalized-cumulative-current}. Lines show the across-trial
mean, and shading denotes the standard error of the mean,
$\operatorname{SD}_i[\mathcal J_i(t)]/\sqrt{N}$, where $N$ is the number of trials.
The reported SNR was calculated from the final current using
Eq.~\ref{eq:trial-current-snr}.

\subsection*{Comparison with alternative dimensionality-reduction methods}
\begin{figure*}[t]
\centering
\includegraphics[width=\textwidth]{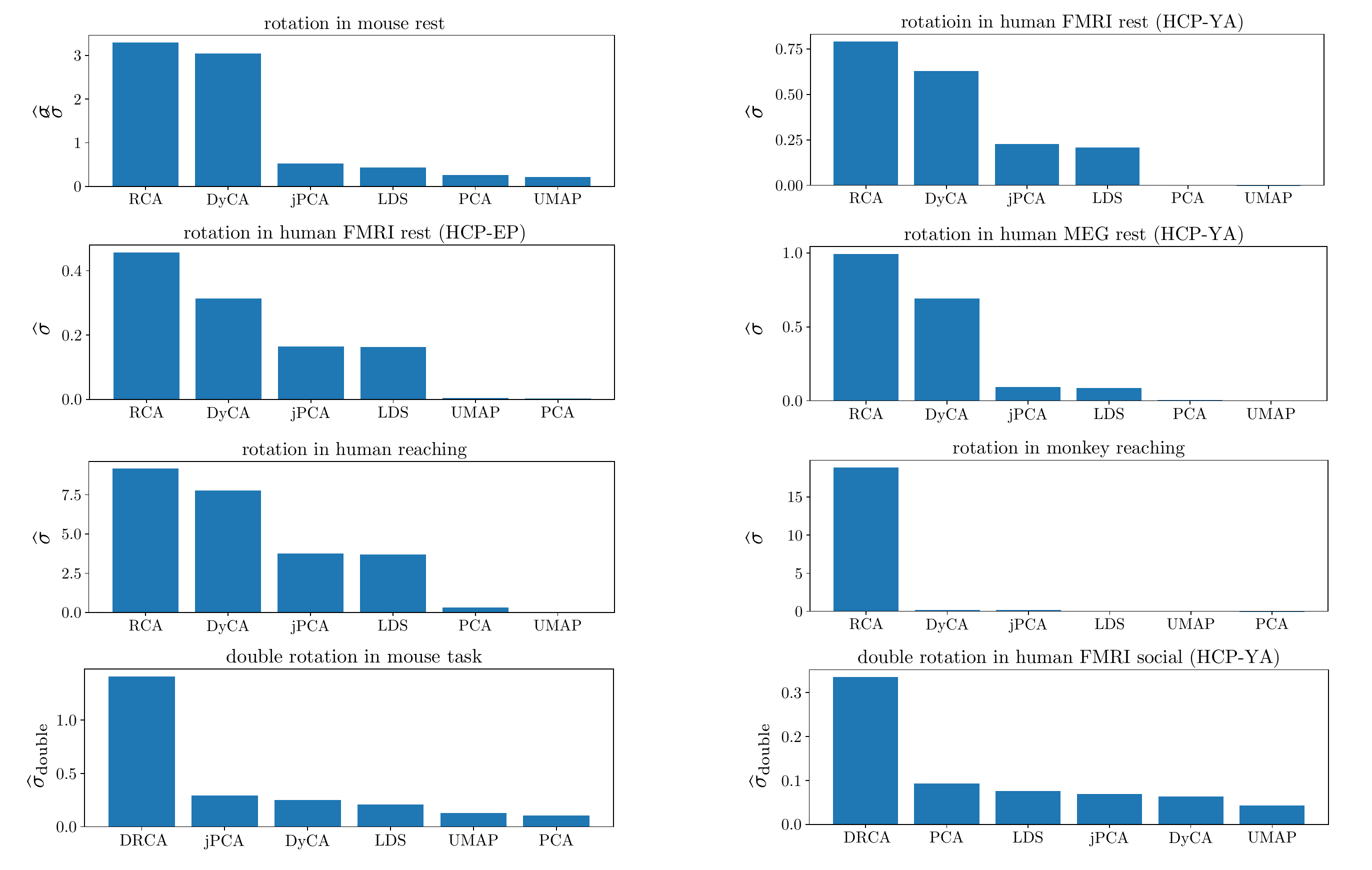}
\caption{\textbf{Comparison of current-based rotation scores across dimensionality-reduction
methods and datasets.} Bars show the surrogate APR $\widehat\sigma$ for
single-rotation analyses (first three rows) and the double-rotation score
$\widehat\sigma_{\mathrm{double}}$ for double-rotation analyses (bottom row). From
top left to bottom right, the datasets are mouse resting-state calcium imaging,
HCP-YA resting-state fMRI, HCP-EP resting-state fMRI, HCP-YA resting-state MEG, human
intracortical activity during reaching, macaque motor-cortical activity during
reaching, mouse calcium imaging during the lever-pull task and HCP-YA fMRI during the
social task. Single-rotation projections were obtained using rotational component
analysis (RCA), dynamical component analysis (DyCA), jPCA, an unconstrained linear
dynamical system (LDS), principal component analysis (PCA) and UMAP. For the
double-rotation analyses, double-rotational component analysis (DRCA) was compared
with the same alternative methods.}
\label{fig:extended-comparison}
\end{figure*}

We evaluated the detected rotational modes by comparing the current-based
scores obtained with RCA, DyCA, jPCA, an unconstrained linear dynamical system (LDS),
PCA and UMAP across the datasets analyzed in this study (Extended Data
Fig.~\ref{fig:extended-comparison}). For the six single-rotation analyses, each method
was evaluated using the surrogate APR $\widehat\sigma$ in
Eq.~\ref{eq:apr-estimator}; for the mouse task and HCP-YA social task, the corresponding
projection was evaluated using the double-rotation score
$\widehat\sigma_{\mathrm{double}}$ in Eq.~\ref{eq:double}. RCA yielded the largest
single-rotation score in every dataset, with DyCA providing the closest alternative in
several cases. This similarity is expected because DyCA favors components whose
derivatives are captured by low-dimensional linear dynamics, and asymmetric linear
coupling between two variables can generate rotation. Its performance was much weaker
for macaque reaching, however, where the rotational trajectories may not be adequately
described by a single linear flow. RCA showed a particularly large advantage in this
dataset, for which the other projections produced little net angular current. DRCA
likewise produced substantially larger double-rotation scores than the comparison
methods in both mouse and human task data.

Because RCA and DRCA directly optimize the respective scores, these comparisons do not
provide an independent test of predictive performance. Nevertheless, $\widehat\sigma$ is a
physically motivated, current-based measure of the strength and consistency of directed
circulation; hence, the comparison shows that RCA and DRCA detect stronger and more consistent rotational dynamics. 

\subsection*{Mouse response-drift model}
Mouse activity was projected onto the RCA plane identified from resting-state data and
aligned to cue onset. We pooled response and nonresponse trials and represented the
instantaneous velocity by the displacement between successive samples. The plane was
divided into a $10\times10$ grid spanning $-2.5\leq x_1,x_2\leq2.5$. Within each grid
cell, the pre-cue drift $\bm f^{\mathrm{pre}}$ was estimated by averaging the
displacements from the ten frames immediately preceding cue onset, whereas the
post-cue drift $\bm f^{\mathrm{post}}$ was estimated from the first five post-cue
frames. Only cells containing at least 1,000 state--displacement pairs in the
corresponding window were retained. Because the calcium signals were sampled at
approximately 30~Hz, the pre-cue and post-cue windows spanned approximately 0.33~s and
0.17~s, respectively.

We fitted the pre-cue field with the linear model
$\bm f^{\mathrm{pre}}(\bm x)=\bm A\bm x+\bm b$ by weighted least squares, using the
number of samples in each cell as the weight. 
The empirical cue-evoked response field was
$\bm f^{\mathrm{rsp}}=\bm f^{\mathrm{post}}-\bm f^{\mathrm{pre}}$ in cells for
which both drift estimates were available. We defined the response direction
$\hat{\bm n}_{\parallel}$ as the direction of the mean response-drift vector across
these cells. The response-drift magnitude was then fitted with 
Eq.~\ref{eq:response-drift}. 

For each trial, we evaluated the modeled post-cue field at the cue-onset state and
defined the scalar drift score
$f^{\mathrm{post}}_{\parallel}=\bm f^{\mathrm{post}}(\bm x^{\mathrm{cue}})\cdot
\hat{\bm n}_{\parallel}$. In Fig.~\ref{fig:model}, data from all fifteen task days were used for fitting. In Fig.~\ref{fig:learning}, to account for changes in the response field during
learning, the parameters of $f^{\rm rsp}$ were estimated
separately in five nonoverlapping three-day training windows (Extended Data
Fig.~\ref{fig:extended-learning}); the pre-cue field and
$\hat{\bm n}_{\parallel}$ were held fixed. For the function
$P^{\rm rsp}(f^{\mathrm{post}}_\parallel)$, scores from all trials were divided into
30 equally spaced bins over their observed range, and bins containing fewer than 50
trials were excluded. The response probability in each retained bin was the fraction
of trials followed by a lever pull, with binomial standard errors. We fitted
Eq.~\ref{Prsp} to these probabilities by nonlinear least squares, weighting each
residual by the square root of the number of trials in that bin. 

\subsection*{Recurrent neural network model}
We implemented a discrete-time leaky RNN with one scalar input, $N=64$ hidden units
and one scalar output. Its hidden state followed Eq.~\ref{eq:rnn-dynamics}, and the
output was
\begin{equation}
y_t=\bm W_{\mathrm{out}}\bm h_t.
\label{eq:rnn-output}
\end{equation}
We used $\alpha=0.2$ and $\sigma=0.5$, initialized the hidden state to zero at the
beginning of every trial and drew an independent Gaussian noise vector at every time
step. The elements of $\bm W_{\mathrm{in}}$ were drawn from
$\mathcal N(0,1)$, whereas those of $\bm W_{\mathrm{out}}$ and
$\bm W_{\mathrm{rec}}$ were drawn from $\mathcal N(0,1/N)$. The recurrent bias was
initialized from the PyTorch default uniform distribution
$\mathcal U(-N^{-1/2},N^{-1/2})$. The input and output weights were fixed after
initialization; only $\bm W_{\mathrm{rec}}$ and $\bm b$ were trained.

Each training trial comprised 100 time steps. A unit-amplitude cue was presented for
one time step at an onset sampled uniformly from steps 10--79. The response window
comprised the following $W=10$ steps, with no intervening delay. Rather than training
the network to match a prescribed output trajectory, we trained it to maximize the net
output displacement over this window. The sum of the one-step output increments
telescopes, so for a batch of $B$ trials with cue onsets $t_i$, the loss can be written
in endpoint form as
\begin{equation}
\mathcal L=-\frac{1}{BW\lVert\bm W_{\mathrm{out}}\rVert}
\sum_{i=1}^{B}\left(y_{t_i+W}^{(i)}-y_{t_i}^{(i)}\right).
\label{eq:rnn-loss}
\end{equation}
The normalization by $W$ expresses the endpoint displacement per time step. Because
$W$ was fixed, however, the loss rewarded net displacement at the end of the response
window and did not independently constrain response speed or the shape of the output
trajectory within the window.
We minimized this loss by backpropagation through time using Adam with a learning rate
of $10^{-3}$, a batch size of 64 and 1,000 parameter-update iterations. For subsequent
analyses, the response strength of trial $i$ was defined as
$v_i^{\mathrm{rsp}}=(y_{t_i+W}^{(i)}-y_{t_i}^{(i)})/
[W\lVert\bm W_{\mathrm{out}}\rVert]$, corresponding to the single-trial contribution
to Eq.~\ref{eq:rnn-loss} without the leading minus sign.

To identify spontaneous rotational dynamics, we simulated the trained network with
zero input. We generated 200 trajectories, discarded the first 100
steps of each as burn-in and retained the following 500 steps. RCA was fitted to these
zero-input trajectories without access to the cue or response
strength. 

For the drift fields in Fig.~\ref{fig:model}H,I, we simulated 100,000 noisy trials
after a 150-step burn-in and sampled cue onset uniformly over the following 60 steps.
We estimated the pre-cue and post-cue drifts from mean one-step displacements within
five-step pre-cue and post-cue windows, respectively. The response direction $\hat{\bm n}_{\parallel}$ was defined as the normalized
spatial mean of
$\bm f^{\mathrm{rsp}}=\bm f^{\mathrm{post}}-\bm f^{\mathrm{pre}}$ over cells in
which both fields were available.

The response-strength map in Fig.~\ref{fig:model}J was estimated from a separate set of
6,000 noisy random-cue trials. Cue-onset state was defined as the state one step before
the cue, and grid cells containing fewer than 15 trials were masked. For
Fig.~\ref{fig:model}K, we simulated 3,000 trials with a fixed cue after burn-in,
projected displacement from the cue-onset state onto
$\hat{\bm n}_{\parallel}$ and divided trials at the median response strength. For
Fig.~\ref{fig:model}L, each trial was assigned the value of
$f^{\mathrm{post}}_{\parallel}=\bm f^{\mathrm{post}}\cdot
\hat{\bm n}_{\parallel}$ in its cue-onset grid cell.

To construct the RNN analogue of Fig.~\ref{fig:mouse}H, we simulated an independent
set of 20,000 random-cue trials and retained the 64-dimensional hidden state one step
before cue onset together with the response strength $v_i^{\mathrm{rsp}}$. The RCA,
PCA and jPCA planes were learned from the same zero-input trajectories used to
characterize spontaneous network activity; jPCA was fitted in the leading six-PC
subspace. We additionally generated 1,000 random orthonormal planes by QR
decomposition of Gaussian random matrices. For every plane, cue-onset states were
centered, oriented with the first coordinate along the leading within-plane covariance
axis and scaled to unit variance along each coordinate. We divided the resulting plane
into a $15\times15$ grid spanning $-2.5\leq x_1,x_2\leq2.5$ and excluded cells with
fewer than 50 trials. The mean response-strength map was fitted with the same
radial-plus-directional form as Eq.~\ref{eq:directional-gradient}, replacing response
probability with $v^{\mathrm{rsp}}$. 

\begin{figure}[t]
\centering
\includegraphics[width=0.5\textwidth]{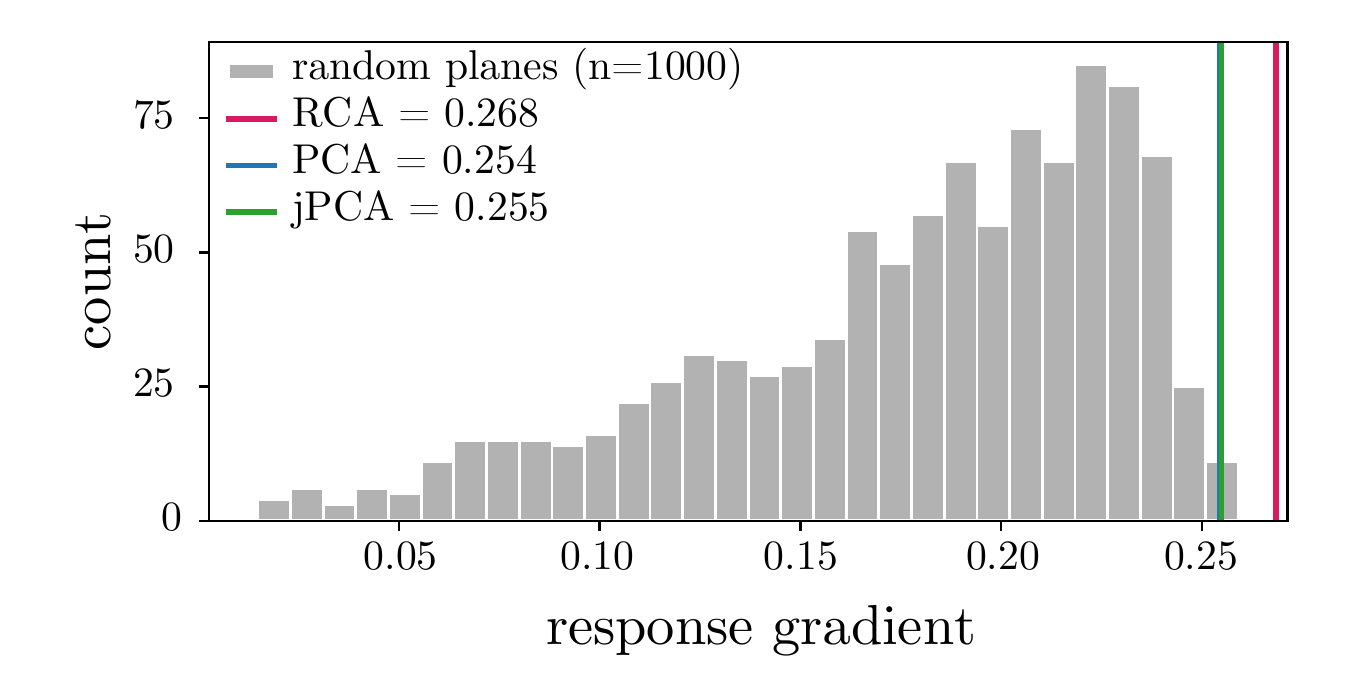}
\caption{\textbf{The spontaneous RNN rotational plane is strongly aligned with the
cue-evoked response.} Distribution of response-gradient magnitudes obtained from
1,000 random orthonormal planes in the RNN hidden state. Vertical lines indicate the
gradients obtained in the RCA (magenta; 0.268), PCA (blue; 0.254) and jPCA (green;
0.255) planes. Projection planes were learned from zero-input network activity, without
access to the cue or response strength, and all projected coordinates were scaled to
unit variance before fitting. The RCA gradient exceeded the gradients from all random
planes.}
\label{fig:extended-rnn}
\end{figure}

The RCA plane produced a gradient of 0.268 and exceeded all 1,000 random-plane
values (Extended Data Fig.~\ref{fig:extended-rnn}). However, it is only slightly
larger than those of PCA (0.254) and jPCA (0.255). This is probably because the RNN is itself low-dimensional as it is trained from a simple task. Therefore, different dimensional reduction techniques may converge on the same plane. 

\subsection*{Held-out prediction from rotational states and pupil-indexed arousal}
For the receiver operating characteristic analysis in Fig.~\ref{fig:double}D, we
randomly divided recording sessions into training (75\%) and test (25\%) sets. The RCA and
DRCA projection vectors were fixed before this supervised analysis. Using only the
training sessions, we separately fitted the response-probability model
$P(r,\theta)=P_0+[f_0+c\sin(\theta-\theta_0)]r$ to the cue-onset RCA state and to the
DRCA state 80 frames (2.7~s) pre-cue. Each fitted function assigned a scalar
response score to every trial.

We then fitted three logistic-regression classifiers using the training sessions: one
based on the RCA score, one based on the DRCA score and one based on both scores.
Predictors were standardized using the training data, and the classifiers used
$L_2$ regularization with $C=1$. The response-probability fits, feature scaling and
classifier parameters were applied without refitting to trials from the held-out test
sessions. The receiver operating characteristic curves and AUC values in
Fig.~\ref{fig:double}D were calculated exclusively from these held-out predictions.

For the comparison with pupil-indexed arousal (Extended Data
Fig.~\ref{fig:extended-pupil}), we used the continuous right-eye pupil-diameter trace
$d(t)$, measured in pixels. Pupil diameter was converted separately for each session to
a robust arousal score,
\begin{equation}
z_{\mathrm{arousal}}(t)=
\frac{d(t)-\operatorname{median}[d]}
{1.4826\,\operatorname{median}\!\left[\left|d-\operatorname{median}[d]\right|\right]},
\end{equation}
where the median and median absolute deviation were calculated over all finite,
positive pupil-diameter samples in that session; the factor 1.4826 makes the scaled
median absolute deviation consistent with the standard deviation for Gaussian
data~\cite{rousseeuw1993alternatives}.
Nonfinite or nonpositive samples, as well as samples with
$|z_{\mathrm{arousal}}|>20$, were treated as tracking artifacts and set to missing. For
each trial, we used the arousal score from the nearest time point with a finite value,
without temporal interpolation. We retained trials with finite arousal values and finite
cue-onset RCA coordinates and applied the same 75:25 session-level splitting procedure
described above. This yielded 35,547 training trials from 267 sessions and 12,272 test
trials from 89 held-out sessions. Using only the training sessions, we fitted the RCA
response score with the radial-sine model above. We also divided pupil arousal into 20
equally spaced bins, retained bins containing at least 25 trials and fitted the binned
response probabilities with a cubic polynomial weighted by the square root of the bin
count. Predicted probabilities were clipped to the interval $[0,1]$.

The RCA-only model was an $L_2$-regularized logistic regression using the fitted RCA
response score. The arousal-only model used the cubic fit directly, and the combined
model was an $L_2$-regularized logistic regression using both the RCA score and the
cubic arousal prediction. Logistic-regression predictors were standardized using the
training sessions only. The receiver operating characteristic curves and AUCs in
Extended Data Fig.~\ref{fig:extended-pupil}B were evaluated exclusively on the held-out
sessions. Extended Data Fig.~\ref{fig:extended-pupil}A is a descriptive visualization:
the arousal bins and cubic curve shown there were refitted using all matched trials and
were not used to calculate the held-out AUCs.

\subsection*{Simulation of the learning model}
The phenomenological learning simulation used the affine pre-cue drift
$\bm f^{\mathrm{pre}}$, global response direction $\hat{\bm n}_{\parallel}$ and
response-probability function in Eq.~\ref{Prsp} inferred from the mouse data. We
approximated the distribution of observed cue-onset states by fitting an isotropic Gaussian
$\mathcal{N}(\bm\mu,\sigma_x^2\bm I)$. For the illustration in
Fig.~\ref{fig:learning}F, we drew 150 states from this distribution, evaluated their
intrinsic drift and assigned each state a binary response outcome with probability
$P^{\mathrm{rsp}}(\bm f^{\mathrm{pre}}(\bm x)\cdot
\hat{\bm n}_{\parallel})$. The displayed response and nonresponse groups were randomly
subsampled to equal sizes, whereas the Fisher linear discriminant shown in the panel
was fitted using all 150 samples.

To simulate learning, we drew a fixed ensemble of $M=200$ cue-onset states from the
same Gaussian distribution and initialized
$\bm f^{\mathrm{rsp}}_0=\bm 0$ and $\bm k=\bm 0$. At each learning iteration, we
evaluated Eq.~\ref{eq:learning-field} at every sampled state $\bm x_i$ and computed
\begin{equation}
\bm f_i^{\mathrm{post}}=\bm f^{\mathrm{pre}}(\bm x_i)
+\bm f^{\mathrm{rsp}}(\bm x_i),\qquad
p_i=P^{\mathrm{rsp}}\!\left(\bm f_i^{\mathrm{post}}\!\cdot\!
\hat{\bm n}_{\parallel}\right).
\end{equation}
Rather than drawing a single response outcome at every iteration, we used the
corresponding mean-field update:
\begin{equation}
\begin{aligned}
\bm f^{\mathrm{rsp}}_0&\leftarrow(1-a_1)\bm f^{\mathrm{rsp}}_0
+b_1\langle\bm f^{\mathrm{post}}\rangle_{\mathrm{rsp}},\\
\bm k&\leftarrow(1-a_2)\bm k+b_2\langle\bm x\rangle_{\mathrm{rsp}}, 
\end{aligned}
\end{equation}
where $\langle\cdot\rangle$ indicates an ensemble average. 
Thus, states and drifts were reinforced in proportion to their probability of producing
a response. 

We ran 50,000 learning iterations with $a_1=0.0014$, $a_2=0.018$,
$b_1=0.001$ and $b_2=0.01$. At the end of the simulation,
$\bm f^{\mathrm{rsp}}_0=(0.0510,-0.0444)$ and
$\bm k=(0.214,0.235)$, and the mean modeled response probability across the sampled
states was 0.584. The resulting response-drift field is shown in
Fig.~\ref{fig:learning}G. 

\clearpage
\section*{Data availability}
The BraiDyn-BC calcium-imaging dataset is described by Kondo et al.~\cite{kondo2025multimodal}.
HCP-YA fMRI and MEG data were obtained from the Human Connectome Project, subject to its
data-use terms. The derived AAL116-parcellated HCP-EP time series were obtained from the
archive accompanying Hancock et al.~\cite{hancock2023metastability}. The human intracortical reaching data are described by Tian et
al.~\cite{tian2026firing}; the data will be released
upon publication. The monkey reaching dataset is available with the original
study~\cite{churchland2012neural}. 

\section*{Code availability}
Code will be available upon publication. 

\bibliography{references}

\section*{Acknowledgements}
We thank Luca Mazzucato for fruitful discussions. D.E.G. acknowledges support by the NSF-Simons National Institute for Theory and Mathematics in Biology (NITMB) Fellowship supported via grants from the NSF (DMS-2235451) and Simons Foundation (MPS-NITMB-00005320). V.V. acknowledges partial support from the Army Research Office under grants W911NF-22-2-0109 and W911NF-23-1-0212. B.D. is supported by the National Institutes of Health (NIH) grants R01EY037119, R01NS133598, T90DA059109; National Science Foundation (NSF) grant 2605555, and the Simons Foundation Collaboration on the Global Brain. Y.X. is supported by a NeuroData Discovery Grant from the Kavli Foundation.
M.F. and V.V. acknowledge partial support from the France Chicago center through a FACCTS grant. 

We thank Charles Greenspon and John Downey for access to the human participant datasets. We thank Michael Boninger, MD, for serving as the sponsor-investigator of the BCI clinical trial conducted under an FDA Investigational Device Exemption.

This research was partly supported by the National Science Foundation through the Center for Living Systems (grant no. 2317138), the National Institute for Theory and Mathematics in Biology, the Simons Foundation and the Chan Zuckerberg Foundation. This work was completed in part with resources provided by the University of Chicago’s Research Computing Center. We acknowledge the use of ChatGPT (OpenAI) and Claude (Anthropic) for assistance with code development, data-analysis workflows, and language editing. The authors reviewed and verified all AI-assisted outputs and take full responsibility for the content of the manuscript.

\end{document}